\documentclass{emulateapj}

\shorttitle{A Systematic Search for Periodic Quasars in PS1 MD09}
\shortauthors{T. Liu et al.}

\usepackage{epsfig}
\usepackage{natbib}
\usepackage{aas_macros}
\usepackage{amsmath}
\usepackage{epstopdf}
\usepackage{rotating}

\begin{document}

\title{A Systematic Search for Periodically Varying Quasars in Pan-STARRS1: \\ An Extended Baseline Test in Medium Deep Survey Field MD09}
\author{T. Liu \altaffilmark{1,2},  S. Gezari\altaffilmark{1}, W. Burgett\altaffilmark{3}, K. Chambers\altaffilmark{4}, P. Draper\altaffilmark{5}, K. Hodapp\altaffilmark{4}, M. Huber\altaffilmark{4}, R.-P. Kudritzki\altaffilmark{4}, E. Magnier\altaffilmark{4}, N. Metcalfe\altaffilmark{5}, J. Tonry\altaffilmark{4}, R. Wainscoat\altaffilmark{4}, and C. Waters\altaffilmark{4} }
\altaffiltext{1}{Department of Astronomy, University of Maryland, College Park, MD 20742, USA}
\altaffiltext{2}{tingting@astro.umd.edu}
\altaffiltext{3}{GMTO Corp, 465 N. Halstead St, Suite 250, Pasadena, CA 91107, USA}
\altaffiltext{4}{Institute for Astronomy, University of Hawaii at Manoa, 2680 Woodlawn Drive, Honolulu, HI 96822, USA}
\altaffiltext{5}{Department of Physics, University of Durham, South Road, Durham DH1 3LE, UK}


\begin{abstract}

We present a systematic search for periodically varying quasars and supermassive black hole binary (SMBHB) candidates in the Pan-STARRS1 (PS1) Medium Deep Survey's MD09 field. From a color-selected sample of 670 quasars extracted from a multi-band deep-stack catalog of point sources, we locally select variable quasars and look for coherent periods with the Lomb--Scargle periodogram. Three candidates from our sample demonstrate strong variability for more than $\sim$ 3 cycles, and their PS1 light curves are well fitted to sinusoidal functions. We test the persistence of the candidates' apparent periodic variations detected during the $4.2$ years of the PS1 survey with archival photometric data from the SDSS Stripe 82 survey or new monitoring with the Large Monolithic Imager at the Discovery Channel Telescope. None of the three periodic candidates (including PSO J334.2028+1.4075) remain persistent over the extended baseline of $7 - 14$ years, corresponding to a detection rate of $<$ 1 in 670 quasars in a search area of $\approx$ 5 deg$^2$. Even though SMBHBs should be a common product of the hierarchal growth of galaxies, and periodic variability in SMBHBs has been theoretically predicted, a systematic search for such signatures in a large optical survey is strongly limited by its temporal baseline and the ``red noise'' associated with normal quasar variability. We show that follow-up long-term monitoring ($\gtrsim 5$ cycles) is crucial to our search for these systems.

\end{abstract}



\section{Introduction}\label{sec:intro}

Supermassive black holes (SMBHs) appear to be at the centers of most, perhaps all, massive galaxies (e.g. \citealt{Kormendy1995}).  Thus, when two massive galaxies merge in the $\Lambda$CDM Universe, it is expected that their nuclei will form a supermassive black hole binary (SMBHB; e.g. \citealt{Springel2005}). As the binary coalesces, the early stage of its orbital decay is driven by exchanging angular momentum with the circumbinary gas disk through viscosity; at smaller separations ($a < 1$ pc), its orbital decay becomes more dominated by gravitational wave (GW) radiation (e.g. \citealt{Begelman1980}).

However, sub-parsec separation SMBHBs at cosmological distances are too compact to resolve with current, or even future, telescopes. Indirect searches so far, therefore, have been focused on spectroscopy, looking for offset broad lines that suggest two broad line emission regions, each likely associated with each black hole in the binary system \citep{Boroson2009}, or offset or shifted peak of the broad line region (e.g. \citealt{Dotti2009, Eracleous2012}).

Another observational aspect of SMBHBs, however, was much under-exploited until recently --- their potential optical variability. One of the first sub-parsec SMBHB candidates identified via its variability was OJ287 \citep{Sillanpaa1988}, which showed quasi-periodic optical outbursts at intervals of 12 years, with the physical interpretation of the burst being the secondary black hole passing through the accretion disk of the primary (e.g. \citealt{Lehto1996, Valtonen2008, Valtonen2011}). More recently, another sub-parsec SMBHB candidate, PG 1302-102 \citep{Graham2015Nat}, was discovered by the Catalina Real-time Transient Survey (CRTS; \citealt{Drake2009}). Its $V$-band light curve can be fitted to a sinusoidal function with period of 1,884 days and amplitude of 0.14 mag. A physical interpretation of PG 1302-102's periodic variability is relativistic Doppler boosting \citep{DOrazio2015Nat}: in this scenario, where the luminosity is dominated by the steadily accreting secondary black hole and the system is viewed at a high inclination angle, emission from the minidisk of the secondary is Doppler-boosted as the black hole orbits at a moderately relativistic speed (along the line of sight).

Another possible scenario that could give rise to periodic variability is modulated mass accretion in the system. Simulations of an SMBHB embedded in a circumbinary disk show that although the binary tidal torque clears and maintains a low gas density cavity at radius $< 2a$ (where $a$ is the binary separation), materials can penetrate the cavity through a pair of streams and be accreted onto the binary. These simulations have the similar results that for a mass ratio $0.01 \lesssim q \leqslant 1$ --- as expected in the merger of two massive galaxies --- mass accretion through the circumbinary disk is strongly modulated as a result of the binary's orbital motion within the circumbinary disk, including two-dimensional (2D) hydrodynamical \citep{MacFadyen2008}, 3D Newtonian magnetohydrodynamical (MHD) \citep{Shi2012} and Post-Newtonian MHD \citep{Noble2012} for an equal mass binary, and general relativistic (GR) MHD \citep{Gold2014} and 2D hydrodynamical simulations \citep{DOrazio2013} for various mass ratios. In these simulations, the accretion rate varies on a time scale that is on the order of the binary orbital time scale, which is in turn a function of the total black hole mass and orbital separation by virtue of Kepler's law. Assuming that luminosity tracks mass accretion of the circumbinary disk, the former should then vary as the latter varies. For a typical black hole mass of $10^7 M_\odot$ and typical separation $10^3 R_s$, the orbital period is on the order of $\sim$ year, an observationally feasible time scale for current time-domain surveys: $t_{\rm orb} = 0.88\,\mbox{yr} \left(\frac{M}{10^{7} M_\odot}\right) \left(\frac{a}{10^3 R_{s}}\right)^{3/2}$ (where $R_s$ is the Schwarzschild radius: $R_{\rm s} = 2GM/c^2$). 

These theoretically explored variability signatures of an SMBHB, as well as encouraging predictions for the detection rates of periodically varying quasars from SMBHBs in a cosmological context \citep{Haiman2009}, motivated several recent systematic searches in large optical time-domain surveys with a temporal baseline of several years --- \cite{Graham2015Nat} and \cite{Graham2015}, with the CRTS; \cite{Charisi2016}, with Palomar Transient Factory (PTF) and additional data from intermediate-PTF and CRTS; \cite{Zheng2015}, with the Sloan Digital Sky Survey (SDSS) and CRTS; and \cite{Liu2015}, with the Pan-STARRS1 Medium Deep Survey (MDS).

In our pilot study (\citealt{Liu2015}, hereafter L15), we performed a systematic search for SMBHB candidates in MDS's MD09 field and reported our first significant detection of such a candidate, PSO J334.2028+1.4075. As reported in L15, PSO J334.2028+1.4075 has a coherent period of P = 542 $\pm$ 15 days in $g_{\rm P1}$ $r_{\rm P1}$ $i_{\rm P1}$ $z_{\rm P1}$ filters, corresponding to almost 3 cycles of variation that is well fitted to a sinusoidal function. It also has an archival $V$-band light curve from CRTS \citep{Drake2009}. Even though the photometric precisions are not comparable, the CRTS light curve is consistent (in the residual sense) with the PS1 only (PV1) sinusoidal fit over $\sim$ 9 years, or $\sim$ 6 cycles. It is also a radio loud quasar (R = $\log{(f_{\rm 5 GHz}/f_{2500\AA})}$ = 2.30; \citealt{Becker2001}) from the VLA FIRST catalog (FIRST J221648.6+012427; \citealt{White1997}).

\begin{figure}[h]
\centering
\epsfig{file=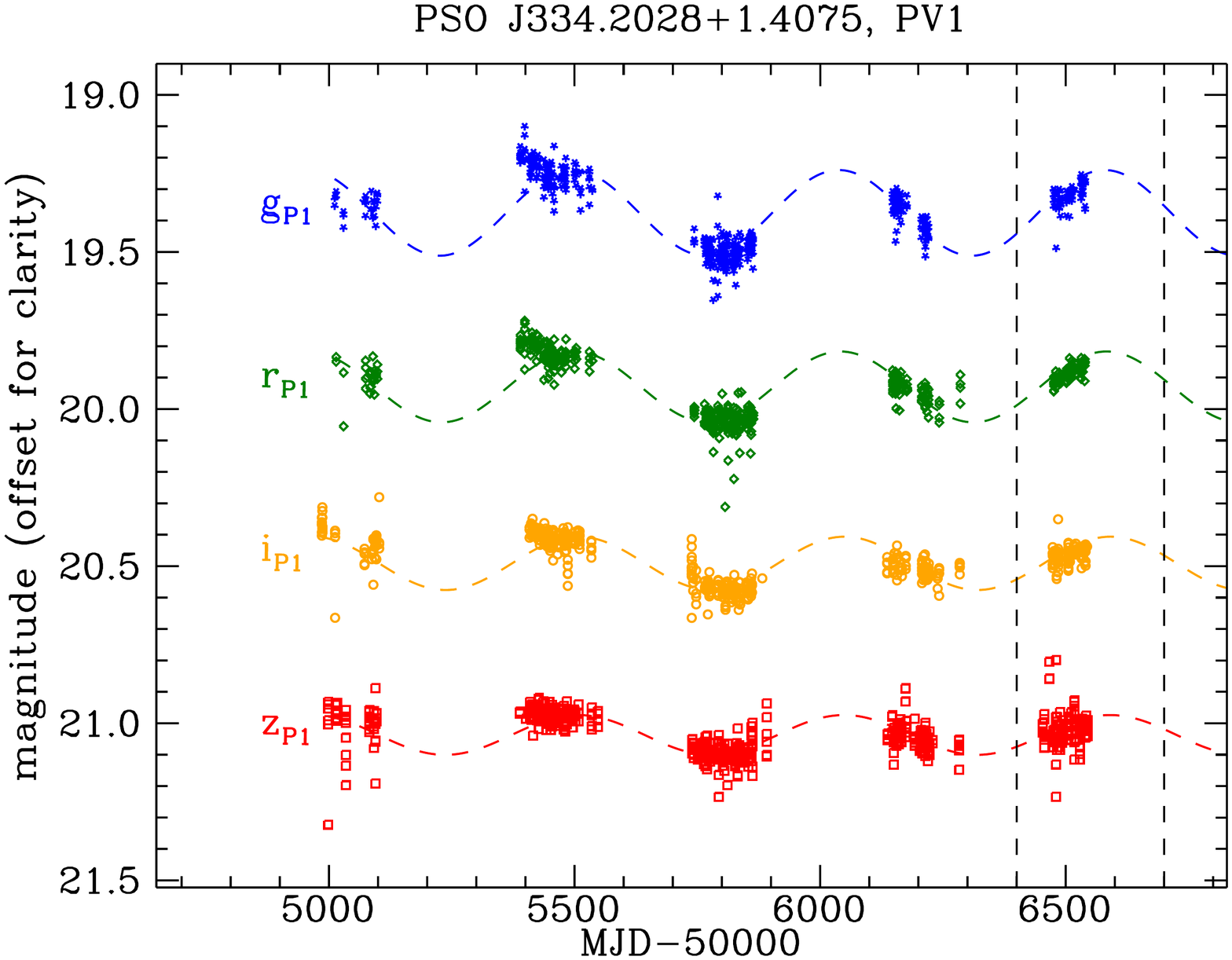,width=0.41\textwidth,clip=}
\epsfig{file=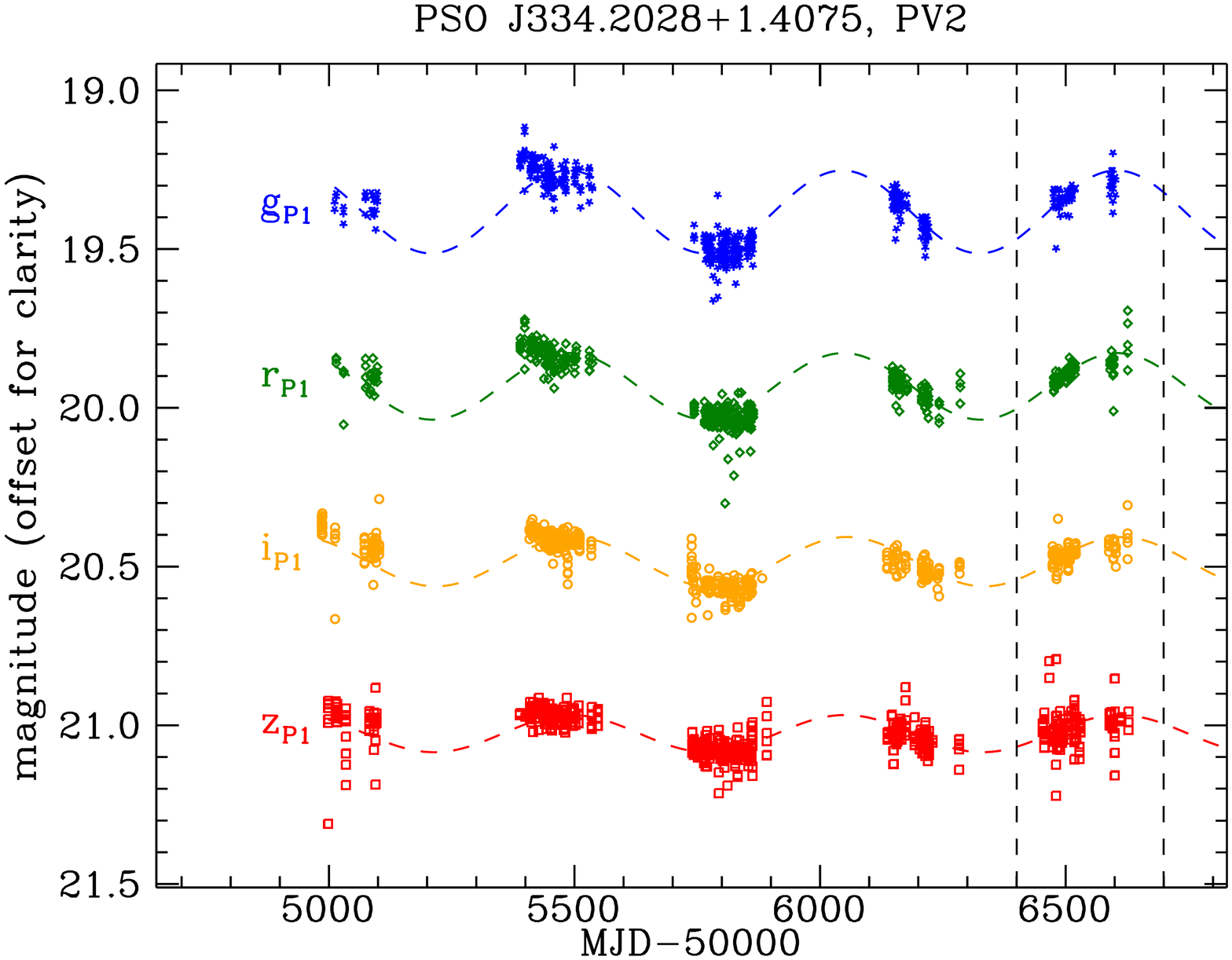,width=0.41\textwidth,clip=}
\caption{In L15, we analyzed the periodic quasar candidate PSO J334.2028+1.4075 based on its light curves in PV1 (upper panel), while its analysis in this paper is based on its light curves from PV2 (lower panel). We note the extra data from the last phase of PS1 MDS are included in PV2 (dashed box), while our conclusions from our new analysis on its significance as a periodic quasar candidate did not change. 4.5 $\sigma$ outliers in $g$ and $z$ filters in both versions have been clipped. The dashed lines are a sinusoid of $P=558$ days (see text for details).}
\label{fig:lett_cand}
\end{figure}

Since then, we have repeated our analysis of MD09 with data Processing Version 2 (PV2) which was made available late-2014 and includes extra data from the final phase of the PS1 survey (Fig. \ref{fig:lett_cand}). We find three periodic quasar candidates that satisfy our selection criteria: a coherent period in at least three filters, an S/N for a sinusoidal fit of $>$ 3 in at least one filter, and a variation over at least 1.5 cycles.  In addition, we use extended baseline data (from archival and new monitoring observations) to test the persistence of our periodic candidates over $5 - 12$ cycles. Recently, it has been pointed out by \cite{Vaughan2016} that the intrinsic red noise (increasing power at lower frequencies) characteristic of quasar variability can easily mimic periodic variability over a small number of cycles, and they emphasize the importance of demonstrating persistence of periodicity over $\gtrsim$ 5 cycles.

This paper thus presents our detailed analysis with MD09 PV2 and is organized as follows.
In \S\ref{sec:mds} we introduce the time domain data set used in this study: MD09 from the Pan-STARRS1 MDS. 
In \S\ref{sec:methods} we describe our methods of variability selection and periodicity search; we also discuss our biases in selecting variable active galactic nuclei (AGNs) in a flux-limited survey like PS1 MDS.
In \S\ref{sec:ext}, we test the persistence of the candidates' periodicity with archival light curves and follow-up imaging. 
In \S\ref{sec:mass}, we measure the black hole mass of binary candidates and calculate their inferred binary parameters.
Finally, in \S\ref{sec:conclude}, we conclude with implications for searches for periodic quasars in a large time-domain survey.
Throughout this paper, we adopt cosmological parameters for a flat universe: $\Omega_{\rm m}$ = 0.3, $\Omega_{\rm \lambda}$ = 0.7, $H_{\rm 0}$ = 70 km s$^{-1}$ Mpc$^{-1}$.


\section{The Pan-STARRS1 Medium Deep Survey}\label{sec:mds}

Pan-STARRS \citep{Kaiser2010} is a multi- filter imaging system designed for sky surveys on a $1.8$ m telescope on the summit of Haleakala in Maui, Hawaii, with a $1.4$ gigapixel camera and a $7$ deg$^2$ field of view. The Pan-STARRS1 (PS1) telescope is operated by the Institute for Astronomy (IfA) at the University of Hawaii and completed its $4.2$ years of operation in the spring of 2014. $\sim$ 25\% of the PS1 telescope time was spent on the MDS, a deep, time domain survey of $10$ circular fields distributed across the sky, totaling $\sim$ $80$ deg$^2$, chosen for their overlap with extragalactic legacy survey fields that have multi-wavelength corollary data. The PS1 MDS cadence typically cycles through the $g_{\rm P1}$, $r_{\rm P1}$, $i_{\rm P1}$ and $z_{\rm P1}$ bands every three nights during the $6$--$8$ months when the field is visible, observing in $g_{\rm P1}$ and $r_{\rm P1}$ on the same night and in the $y_{\rm P1}$ band close to the full Moon (though $y$ band observations were not used in our study due to the poorer sampling). Nightly observations consist of eight 113 s ($g_{\rm P1}$ $r_{\rm P1}$) or 240 s ($i_{\rm P1}$ $z_{\rm P1}$ $y_{\rm P1}$) exposures \citep{Tonry2012MDS}; over the course of MDS, each object is observed $\sim$ 300 times to a 5$\sigma$ (i.e. where $\Sigma$ = 0.217 mag in Figure \ref{fig:phot_err}) limiting magnitude of $\sim$ 22.5 mag in $g_{\rm P1}$ $r_{\rm P1}$ $i_{\rm P1}$ and $\sim$ 22.0 mag in $z_{\rm P1}$ in a single exposure. Individual exposures can be combined into nightly stacks or full-survey-depth ``deep'' stacks, to reach much deeper limits of $\sim 23.5$ mag and $\sim 25$ mag, respectively (in the g$_{\rm P1}$, r$_{\rm P1}$, and i$_{\rm P1}$ bands).

The PS1 photometric calibration includes a combination of ``absolute'' calibration, which translates the number of photons detected to the physical unit of magnitude, and ``relative'' calibration, which removes variations due to the telescope system and atmosphere over the course of the survey. The PS1 absolute photometric calibration is accomplished by observing photometric standard stars from HST's Calspec catalog \citep{Tonry2012photometry}, as part of the Image Processing Pipeline (IPP; \citealt{Magnier2006IPP}). The relative calibration is based on the algorithm of \cite{Padmanabhan2008} which is known as ``Ubercalibration'' (Ubercal). The PS1 Ubercal \citep{Schlafly2012} uses multiple observations of the same non-intrinsically variable sources on photometric nights and demands that the observed magnitude does not change over time and thereby minimizes variations in the zero point. PS1 data are further calibrated through ``Relphot'' \citep{Magnier2013Relphot}, which solves for an additional zero point offset for each exposure, using the Ubercal solutions as a starting point.

In L15, we employed a similar technique adapted from \cite{Bhatti2010} (``ensemble photometry''; \citealt{Honeycutt1992}) which implements the \texttt{ENSEMBLE} \footnote{http://spiff.rit.edu/ensemble/} software package in our attempt to achieve precision photometry with MDS. We constructed an ``ensemble'' of point sources near each target object in a $\sim$0.1 deg$\times$$\sim$0.1 deg field and ran the algorithm iteratively to obtain a least-squares solution that \emph{locally} reduces the scatter for all observations of each source over the course of the survey. However, since the PS1 data products had already been Ubercaled, the overall improvement in our control sample of stars in L15 or our re-analysis with PV2 was not significant enough to justify this time-intensive procedure; thus we do not apply the method of ensemble photometry in the analysis presented in this paper.


\section{Methods}\label{sec:methods}
\subsection{Sample Selection}\label{sec:sample}

We first extracted from the PS1 Science Archive all sources from MD09 that matched the following criteria: 1) they are point sources selected as deep stack mag$_{\rm psf}-$mag$_{\rm Kron}$$<0$ \footnote{Since the Kron radius captures more flux from an extended source than the PSF profile, while for a point source its Kron magnitude should be close to its PSF magnitude.} that have a good point spread function (PSF) quality factor from the IPP (\texttt{psfQF} $> 0.85$), 2) they have $\texttt{stackPSFMag} < 23$ mag, 3) they have at least five detections in each filter, and 4) masks were applied to exclude bad and poor detections (Table \ref{table:flags}). The query resulted in $\sim$ 40,000 point sources, for which we get PSF magnitudes, each with an average of $\sim$ 300 detections in each of the four filters.  

For our color-selection of quasars and stars, we use a catalog of Kron magnitudes extracted from deep stack images in the $g_{\rm P1}$, $r_{\rm P1}$, $i_{\rm P1}$, $z_{\rm P1}$, $y_{\rm P1}$ bands from PS1 MDS as well as in the $u_{\rm CFHT}$ band from the Canada-France-Hawaii Telescope (CFHT) (\citealt{Heinis2016AGN}, hereafter the PS1$\times$CFHT catalog). We use the PS1$\times$CFHT catalog star/galaxy classification, which was determined using a machine learning method of support vector machine (SVM) that was trained on an HST/ACS sample of stars and galaxies and has a completeness of 88.5\% for stars $i_{\rm P1}<$ 21 mag, or 97.4\% of all objects down to $i_{\rm P1}$ = 24.5 mag \citep{Heinis2016SVM}.  We then cross-match (using a 1$\arcsec$ radius) our PS1 Science Archive point sources with point sources in the PS1$\times$CFHT catalog with $r_{\rm P1}<23$ mag (where the star/galaxy separation is the most reliable), and within a 1.59 deg radius from the center of the MD09 field (to avoid edge effects). 

For the $\sim$ 15,000 cross-matched point sources, we then converted their CFHT \footnote{http://www.cadc-ccda.hia-iha.nrc-cnrc.gc.ca/en/megapipe/\\docs/filt.html} $u$ and PS1 $g$ $r$ $i$ $z$ \citep{Tonry2012photometry} band magnitude to the SDSS magnitude system

\begin{eqnarray*}
u_{\rm SDSS} &=& (u_{\rm CFHT} - 0.241\,g_{\rm SDSS})/0.759 \\
g_{\rm SDSS} &=& 0.014 + 0.162\,(g_{\rm P1}-r_{\rm P1}) + g_{\rm P1} \\
r_{\rm SDSS} &=& -0.001 + 0.011\,(g_{\rm P1}-r_{\rm P1}) + r_{\rm P1} \\
i_{\rm SDSS} &=& -0.004 + 0.020\,(g_{\rm P1}-r_{\rm P1}) + i_{\rm P1} \\
z_{\rm SDSS} &=& 0.013 - 0.050\,(g_{\rm P1}-r_{\rm P1}) + z_{\rm P1} \quad, \\
\end{eqnarray*}

\noindent and selected those that have the following quasar colors for their clean separation from stars \citep{Sesar2007}:

\begin{eqnarray*}
u_{\rm SDSS}&-&g_{\rm SDSS} < 0.7 \\
-0.2 < g_{\rm SDSS}&-&r_{\rm SDSS} < 0.5\quad. \\
\end{eqnarray*}

\begin{figure}[h]
\centering
\includegraphics[width=3.5 in]{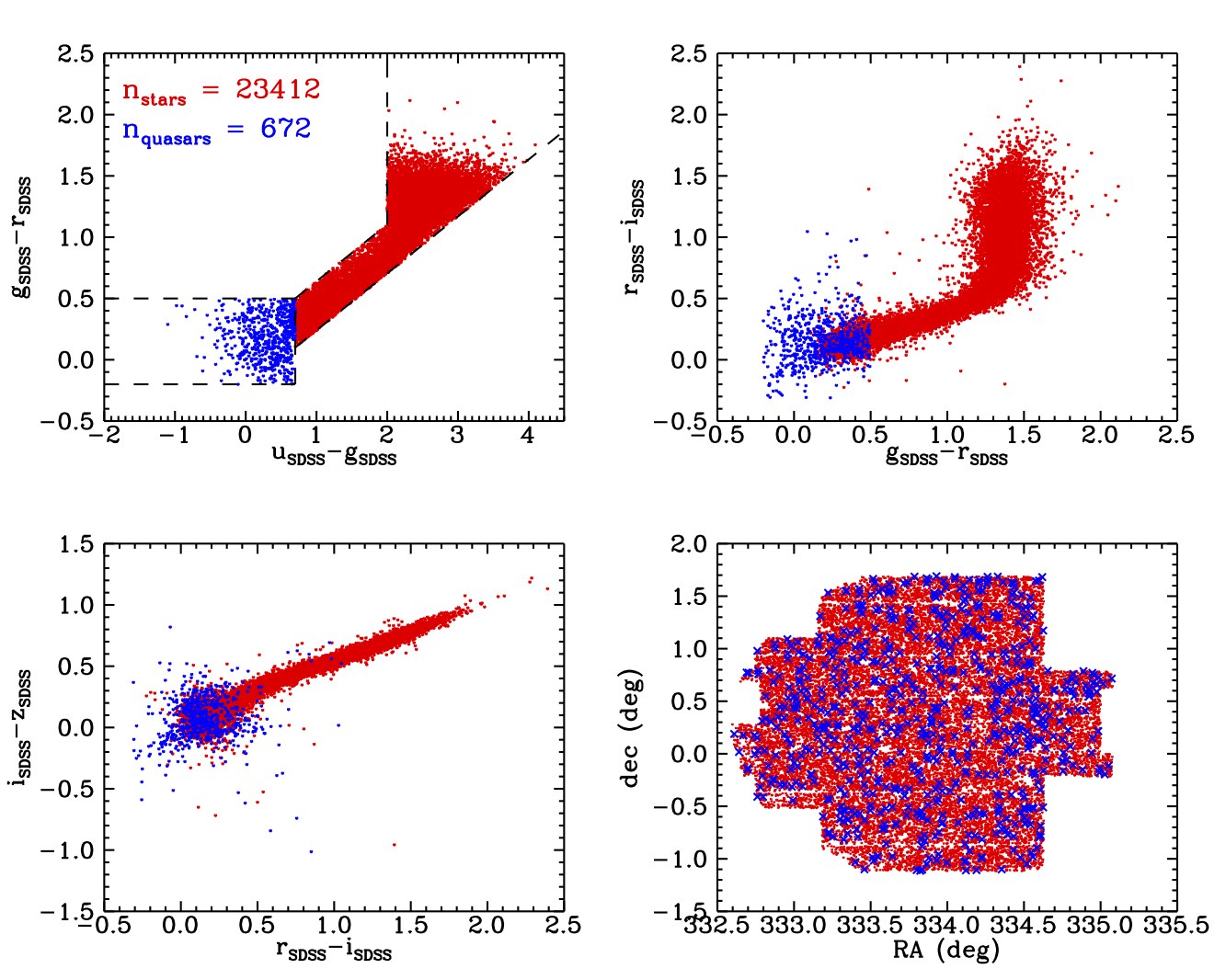}
\caption{First three panels: cross-matched stars and quasars are selected by their SDSS colors (converted from $u_{\rm CFHT}$ $g_{\rm P1}$ $r_{\rm P1}$ $i_{\rm P1}$ $z_{\rm P1}$). In the upper left panel, the regions occupied by quasars (blue) and stars (red) are represented by dashed boxes, and the stellar region does not include RR Lyrae variables.
Bottom right panel: spatial map of all stars and quasars (red dots and blue crosses, respectively) in MD09 that have cross-matches in the PS1$\times$CFHT catalog. The deep stack photometry \citep{Heinis2016AGN} was performed with each PS1 ``sky cell'' as the smallest unit (each MD field is divided into 10$\times$10 such rectangular regions), hence the rectangular shape. The actual search area is smaller than the total area of MD09 field and is about 5 deg$^2$.}
\label{fig:color_diagram}
\end{figure} 

\begin{figure}[h]
\centering
\epsfig{file=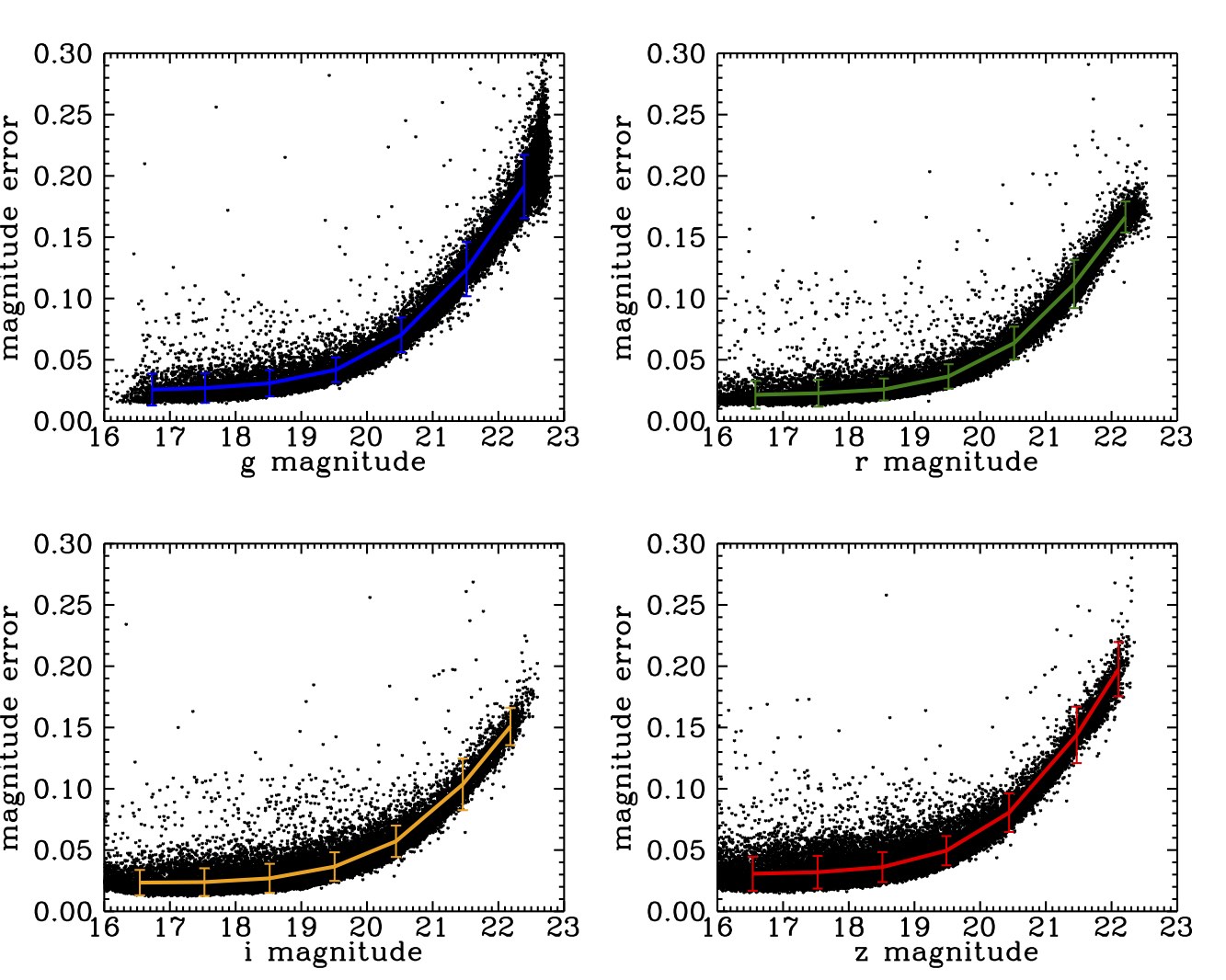,width=0.47\textwidth,clip=}
\caption{For our sample of stars, we plot the observed scatter (standard deviation of the light curve) as a function of magnitude and fit the binned relation to a parabola (Eqn \ref{eqn:sigma_g}--\ref{eqn:sigma_z}). We have masked outliers at the 4.5 $\sigma$ level in our scatter plot, and they are excluded from the binned scatter vs. magnitude relation. The relations are similar in the four filters ($\sim$ 0.03 mag at the bright end and rising to $\sim$ 0.15 mag at the faint end), and the size of the error bars is less than 20 mmag, reflecting the stability of the PS1 system over the course of the survey and zero-point variations mainly due to the atmosphere \citep{Schlafly2012}.}
\label{fig:phot_err}
\end{figure}

\begin{deluxetable}{ll}
\tablecaption{PS1 Quality Flags Used in the Query \label{table:flags}}
\tablehead{
\colhead{Flag Name} & \colhead{Flag Name}
}
\startdata
FAIL & POOR \\
PAIR & SATSTAR \\
BLEND & BADPSF \\
DEFECT & SATURATED \\
CR\_LIMIT & EXT\_LIMIT \\
MOMENTS\_FAILURE & SKY\_FAILURE \\
SKYVAR\_FAILURE & BELOW\_MOMENTS\_SN \\
BLEND\_FIT & SIZE\_SKIPPED \\
PEAK\_ON\_SPIKE & PEAK\_ON\_GHOST \\
PEAK\_OFF\_CHIP & 
\enddata
\end{deluxetable}

We also select stars for our control sample, carefully avoiding the region occupied by RR Lyrae variables (u$_{\rm SDSS}$$-$g$_{\rm SDSS}$ $\sim 1.15$; \citealt{Sesar2007}). The color diagrams of selected quasars and stars are shown in Fig. \ref{fig:color_diagram}. In order to obtain our variability detection limit (\S\ref{sec:var_sel}), we plot their error vs. magnitude relation for our star sample in Fig. \ref{fig:phot_err} and fit the binned relation in each filter to a parabola:

\begin{eqnarray}
\sigma (g) &=& 2.64372 - 0.293112\,g+0.00818841\,g^2 \label{eqn:sigma_g}\\
\sigma (r) &=& 2.39328 - 0.267030\,r+0.00749830\,r^2 \label{eqn:sigma_r}\\
\sigma (i) &=& 2.13028 - 0.237271\,i+0.00666299\,i^2 \label{eqn:sigma_i}\\
\sigma (z) &=& 2.77188 - 0.309921\,z+0.00874017\,z^2 \label{eqn:sigma_z} \quad.\\ \nonumber
\end{eqnarray}


\subsection{Variability Selection}\label{sec:var_sel}

\begin{figure}[h]
\centering
\epsfig{file=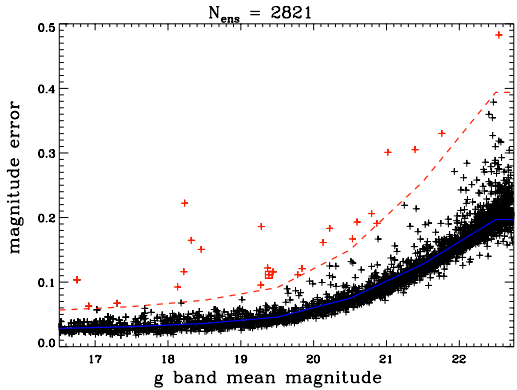,width=0.4\textwidth,clip=}
\caption{Magnitude error (light curve standard deviation) versus magnitude in the $g_{\rm P1}$ filter for all the objects in one ``ensemble'' (crosses). The majority of the 2821 sources in the ensemble are non-variable, and their error vs. magnitude relation (black crosses) can be represented as a piece-wise linear function (blue solid line), and any objects varying above the $2\sigma$ level (red dashed line) are excluded (red crosses). After three iterations, the target quasar of this ensemble (marked with an additional red square) is selected as a variable from this ensemble.}
\label{fig:var_sel}
\end{figure}

To select intrinsic variables from our quasar sample, we perform a variability selection in such way that systematic effects local to the field are minimized: we plot the magnitude error in terms of standard deviation of the light curve as a function of magnitude for each object within $\Delta$R.A. = 0.5 deg and $\Delta$decl. = 0.5 deg from each color-selected quasar. Each of these ``ensembles'' contains $\sim$ 1000 point sources. We calculate the median value for each magnitude bin, while avoiding the bins with less than five stars, and interpolate linearly between the bin centers. The intrinsic variables have a significantly higher magnitude scatter than stars of the same brightness and thus appear as outliers that deviate from the error vs. magnitude trend established by the majority of objects. We iteratively remove variables from the linear interpolation, and after three iterations, those that passed the final 2$\sigma$ detection threshold (have at least twice the magnitude error than the linear interpolation) are tagged as variables (Fig. \ref{fig:var_sel}). Our piece-wise interpolation method is adapted from the variability selection procedure in \texttt{ENSEMBLE} and gives better results than the parabolic fitting method in the previous version that was applied in our analysis in L15.

\begin{figure}[h]
\centering
\epsfig{file=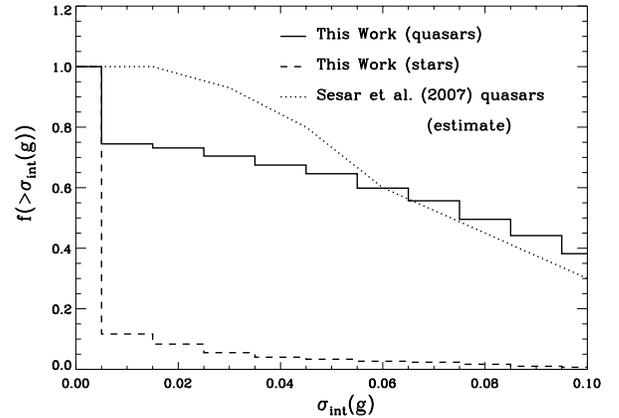,width=0.45\textwidth,clip=}
\caption{The number fraction of our MD09 quasars that vary more than $\sigma_{\rm int}$ (bin size = 0.01 mag) in the $g_{\rm P1}$ band (solid histogram) decreases with increasing intrinsic variability, and is in agreement with results from SDSS Stripe 82 quasars \citep{Sesar2007} for $\sigma_{\rm int} > 0.06$ mag.  The variability fraction of a control sample of stars is shown in the dashed histogram. Plotted with a dotted line are number fractions estimated from \cite{Sesar2007} which were derived from their sample of spectroscopically confirmed quasars.}
\label{fig:var_frac}
\end{figure}

Of the 670 quasars processed through this stage, we flag variables independently in each filter, and further require a variability flag in at least two filters. To compare our quasar sample with previous studies, we calculate their intrinsic variability $\sigma_{\rm int}$ by putting in quadrature the standard deviation of the light curve $\Sigma$ and the photometric error $\sigma$: $\sigma_{\rm int}$ = $\sqrt{\Sigma^2 - \sigma^2}$ \citep{Sesar2007} for $\Sigma>\sigma$ and $\sigma_{\rm int} = 0$ otherwise, where $\sigma$ is the magnitude-dependent photometric error from Eqn. \ref{eqn:sigma_g}--\ref{eqn:sigma_z}. We find the number fraction of quasars varying at the $>\sigma_{\rm int}$ level qualitatively agrees with the results from SDSS Stripe 82 (S82) quasars in \cite{Sesar2007} at $\sigma_{\rm int} > 0.06$ mag, where $\sim 60 \%$ of quasars vary at or above that level, compared to a control sample of stars for which the fraction is $< 5 \%$ (Fig. \ref{fig:var_frac}).  The lower quasar variability fraction that we find for smaller variability amplitudes is likely due to our factor of $\sim 2$ larger photometric errors compared to S82 ($\sigma$(g)$>$0.04 mag vs. $\sigma$(g)$>$0.018 mag).

\pagebreak

\begin{figure}[h]
\centering
\epsfig{file=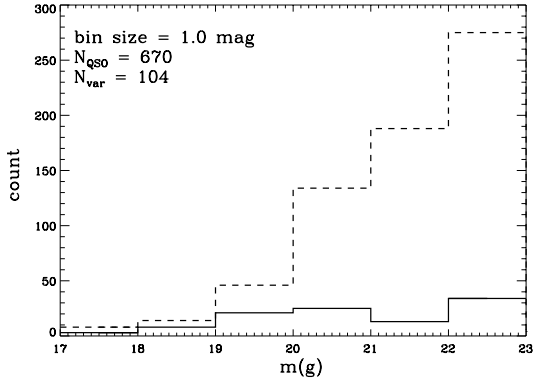,width=0.41\textwidth,clip=}
\caption{The $g_{\rm P1}$ band apparent magnitude distribution of our MD09 quasar sample. The full PS1$\times$CFHT quasar sample (dashed histogram; $N_{\rm QSO}$ = 670) is similar to the distribution derived from the quasar luminosity function (\S\ref{sec:bias}), and our variability selection (solid histogram; $N_{\rm var}$ = 104).}
\label{fig:ps1_dist}
\end{figure} 

\begin{deluxetable*}{lcccccc}
\tablecaption{QLF Model Values Used in Eqn \ref{eqn:phi} -- \ref{eqn:pede} \citep{Ross2013} \label{table:qlf}}
\tablehead{
\colhead{QLF Model} & \colhead{$\alpha$} & \colhead{$\beta$} & \colhead{$M_{i}^{\ast}$} & \colhead{$k_{1}$} & \colhead{$k_{2}$} & \colhead{$\log(\Phi^{\ast})$}
}
\startdata
PLE ($0.3<z<2.2$) & -1.16 & -3.37 & -22.85 & 1.241 & -0.249 & -5.96 \\
LEDE ($2.2<z<3.0$) & -1.29 & -3.51 & -26.57 & -0.689 & -0.809 & -5.93
\enddata
\end{deluxetable*}


\subsection{Selection Bias}\label{sec:bias}

To investigate the possible biases in our variability selection in a flux-limited survey, we simulated $\sim$9000 quasars whose population is derived from the quasar luminosity function (QLF).
The QLF is defined as the number of quasars per co-moving volume per unit magnitude and is described by a broken power law:

\begin{equation}
\Phi(M,z) = \frac{\Phi^{\ast}}{10^{0.4(\alpha+1)[M-M^{\ast}(z)]}+10^{0.4(\beta+1)[M-M^{\ast}(z)]}}\quad,
\label{eqn:phi}
\end{equation}

\noindent where $M_{\ast}$ is the characteristic break absolute magnitude, and $\alpha$ and $\beta$ are the slopes of the QLF at the faint end and bright end, respectively. At lower redshifts ($0.3<z<2.2$), the QLF is described by a pure luminosity evolution (PLE) model, where the characteristic number density $\Phi^{\ast}$ remains constant while $M^{\ast}$ evolves with redshift quadratically (in $i$ band):

\begin{equation}
M^{\ast}_{i}(z) = M^{\ast}_{i}(z=0) - 2.5(k_{1}z+k_{2}z^{2})\quad.
\label{eqn:ple}
\end{equation}

\noindent At higher redshifts ($z>2.2$), it is necessary to model the QLF as the results of luminosity evolution and density evolution (LEDE), where $\Phi^{\ast}$ and $M^{\ast}$ evolve independently with redshift:

\begin{eqnarray}
\log \Phi^{\ast}(z) = \log \Phi^{\ast}(z=2.2)  - c_{1}(z-2.2)\\
M^{\ast}_{i}(z) = M^{\ast}_{i}(z=2.2) - c_{2}(z-2.2)\quad.
\label{eqn:pede}
\end{eqnarray}

\noindent We adopt the values for the constants (Table \ref{table:qlf}) given in \cite{Ross2013}, which expanded the redshift range in previous QLF studies (e.g. \citealt{Richards2006QLF}) to $0.3<z<3.5$.

In each redshift bin of size $0.4$ from $z = 0.3-3.0$, we integrate over the absolute magnitude range ($-14<M_{i}<-30$ mag, $\Delta M_{i} = 1$ mag) and use the cosmology calculator by \cite{Wright2006} to calculate the co-moving volume of each shell from $z$. From this quasar redshift distribution, we then populate each absolute magnitude bin for each redshift according to the $\Phi(M,z)$. To convert absolute magnitudes at different redshifts to the observed frame, it is necessary to apply the $K$ correction: $m = M_{z=2} + {\rm distance\,modulus} + K(z)$, where we adopt the values for $K(z)$ from \cite{Richards2006QLF}. We also converted the absolute magnitude at $z = 2$ in the QLF to $z = 0$: $M_{z=0} = M_{z=2} +0.596$, assuming a constant quasar spectral power law index of $\alpha = -0.5$ \citep{Richards2006QLF}.

With a distribution of quasars in redshifts and absolute magnitudes, we then adopt the empirical relations from \cite{Heinis2016AGN} to calculate the expected variability amplitude given the quasar's (rest frame) absolute magnitude. Using difference imaging of $\sim$1000 variability selected AGNs from PS1 MDS, \cite{Heinis2016AGN} measured the anti-correlation between fractional flux variability and the AGN luminosity in the $g_{\rm P1}$ $r_{\rm P1}$ $i_{\rm P1}$ $z_{\rm P1}$ bands, and here we convert their relations to the magnitude space:

\begin{eqnarray}
\log\Big(\frac{\Delta f}{f}\Big)_{g} &=& 0.17\,M_{g}+3.40 \label{eqn:frac_var1}\\
\log\Big(\frac{\Delta f}{f}\Big)_{r} &=& 0.18\,M_{r}+3.64 \label{eqn:frac_var2}\\
\log\Big(\frac{\Delta f}{f}\Big)_{i} &=& 0.20\,M_{i}+4.02 \label{eqn:frac_var3}\\
\log\Big(\frac{\Delta f}{f}\Big)_{z} &=& 0.20\,M_{z}+4.07 \label{eqn:frac_var4}\quad,
\end{eqnarray}

\noindent where $\log(\Delta f/f)$ was measured from the maximum flux on the difference images over the course of PS1 MDS, and the absolute magnitude $M$ of the AGN was derived from SED fitting \citep{Heinis2016AGN}. 

We calculate the expected fractional flux variability $\log{(\Delta f/f)}$ from Equations \ref{eqn:frac_var1} -- \ref{eqn:frac_var4} and convert to $\Delta m$: $\Delta m  = 2.5\,\log{[1+10^{\log{(\Delta f/f})}]}$.  Finally, we scale up to the expected variability amplitude in its rest frame wavelength using the relation from \cite{VandenBerk2004}: $v(\lambda) = 0.616\,\rm exp(-\lambda_{\rm rest}$/988\AA)+0.164 (where $v$ is variability amplitude measured in terms of ``structure function,'' also in units of magnitude). To estimate the variability detection threshold for simulated quasars, we calculate the expected photometric errors $\sigma$ for a given quasar's apparent magnitudes $m$ in the four PS1 filters from Equations \ref{eqn:sigma_g} -- \ref{eqn:sigma_z}. We adopt average quasar colors of $g_{\rm P1} -r_{\rm P1} = 0.14$ mag, $r_{\rm P1} -i_{\rm P1} = 0.15$ mag, and $i_{\rm P1} -z_{\rm P1} = 0.08$ mag from our sample of PS1$\times$CFHT quasars in Fig. \ref{fig:color_diagram} as a proxy for our color selection. Note that this color box is a valid assumption for quasars at $z<3$ \citep{Richards2002SDSSQSO}. Next, we apply the same magnitude cut as our PS1 quasar sample ($r<$ 23 mag). And finally, we assume that $\Sigma = 0.023+ 0.27\, \Delta m$, the average empirical relation we found for our MD09 variable quasar sample, and using the same variability selection criteria as described in \S\ref{sec:var_sel}, a quasar varying at $\Sigma>2\sigma$ level in any and at least two filters is flagged in our simulation as a variable.

\begin{figure*}[h]
\centering
\includegraphics[width=3.5 in]{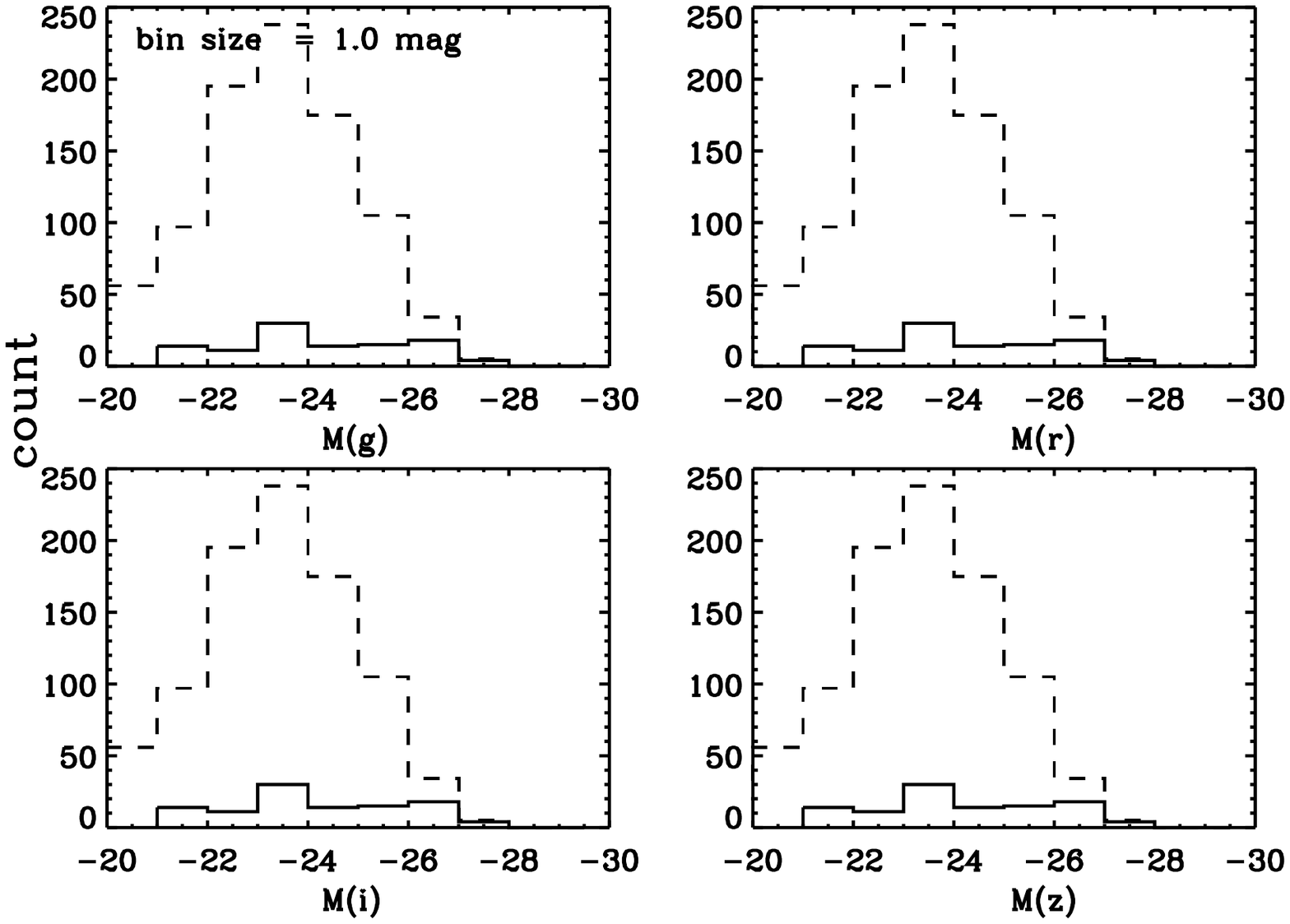}
\includegraphics[width=3.5 in]{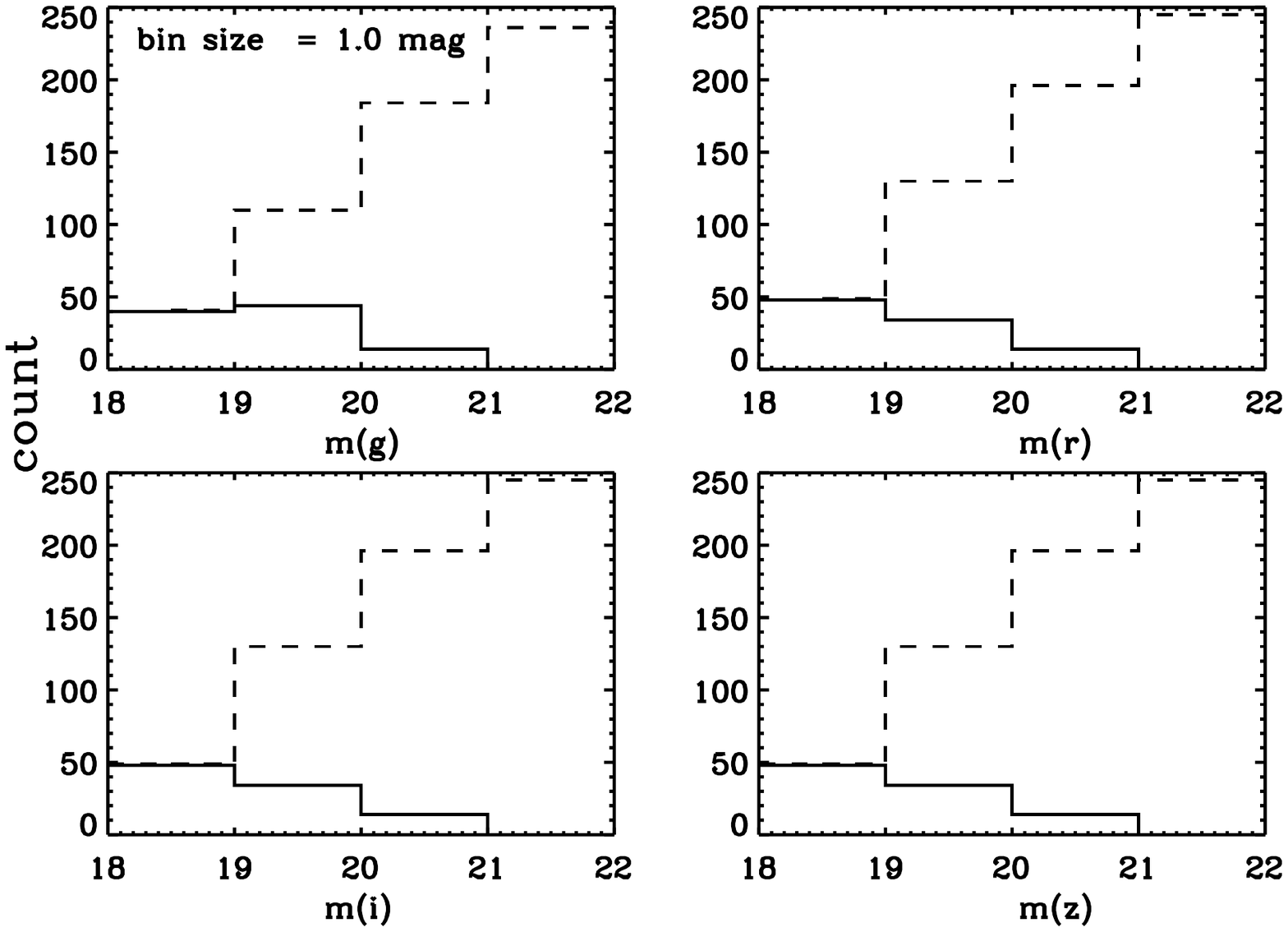}
\includegraphics[width=3.4 in]{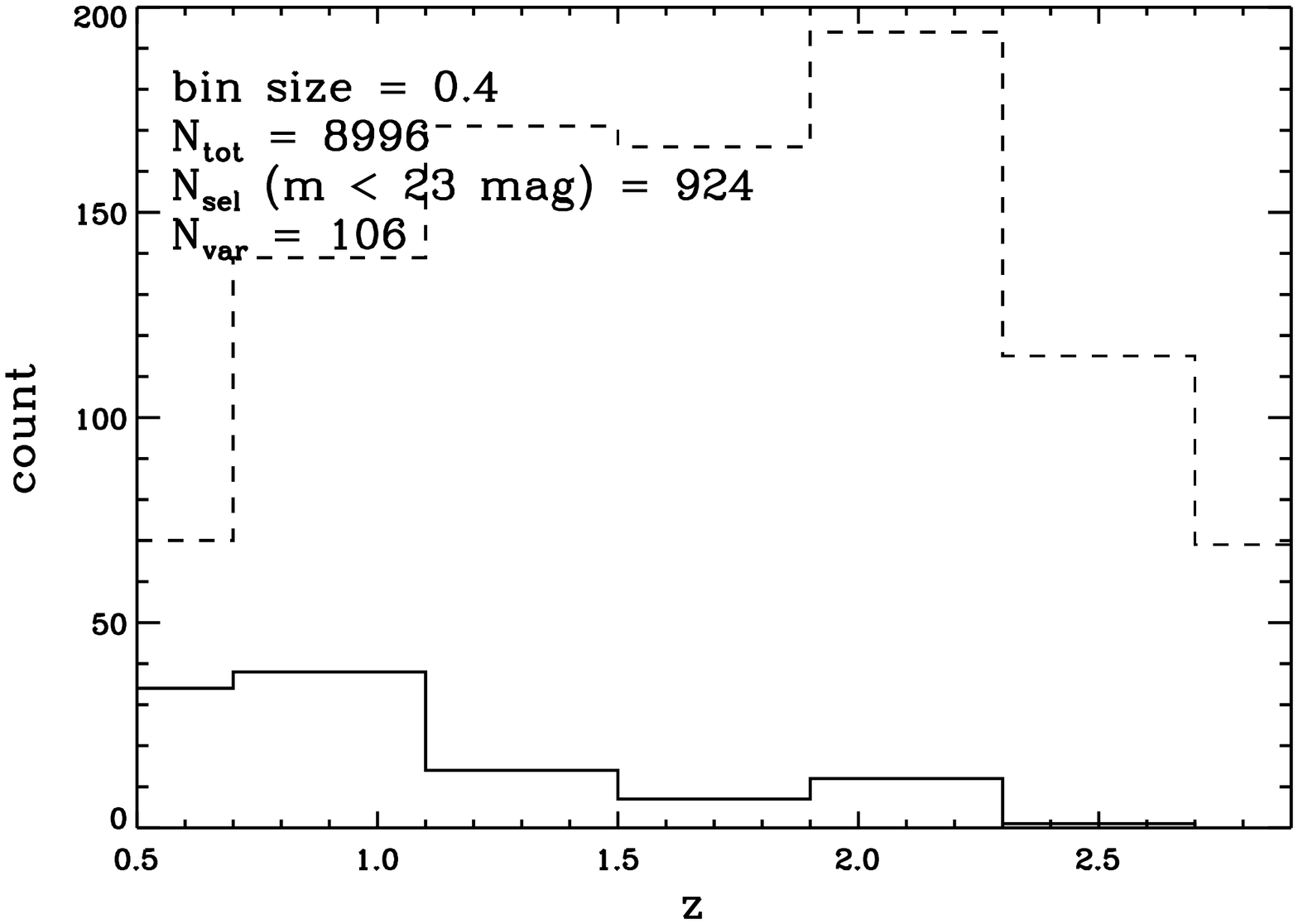}

\caption{To simulate the detectability of quasars by our selection criteria whose variability amplitude obeys the empirical relations of normal AGN variability found in \cite{Heinis2016AGN}, we draw $\sim$9000 quasars from the redshift and absolute magnitude distributions derived from the quasar luminosity function in \cite{Ross2013} (full distribution not included). Among the ``visible'' quasars ($m<23$ mag; dashed histogram), our variability selection is biased toward lower luminosity (and thus in general more variable) quasars at lower redshifts (solid histogram).}
\label{fig:sim_dist}
\end{figure*}

The redshift, absolute magnitude, and apparent magnitude distributions of the ``variable'' simulated quasars are compared with the ``visible'' simulated quasars ($m < 23$ mag) in Fig. \ref{fig:sim_dist}. Among the 924 ``visible'' quasars ($m<$ 23 mag), and assuming all obey the \cite{Heinis2016AGN} relation, 106 (or 11.5\%) are selected as ``variables,'' comparable to the observed variable quasar fraction of 15.5\% we find in the MD09 sample (see Table \ref{table:bythenumbers}), while both the ``visible'' and ``variable'' samples have similar apparent magnitude distributions with MD09 quasars (Fig. \ref{fig:ps1_dist}). We find that in a sample of normal quasars or AGNs, our $2\sigma$ variability selection is biased toward brighter quasars ($m < 21$ mag) at lower redshifts ($z < 1$), with a relatively flat distribution in luminosity ($-21$ mag $ > M > -27$ mag). Understanding this selection bias will be important in calculating the volume density of SMBHB candidates in our final sample.


\subsection{Selecting Periodic Quasar Candidates}\label{sec:periodic}

We then began to search for potential periodic signatures using the Lomb--Scargle (LS) periodogram, a Fourier analysis technique for unevenly spaced data with noise \citep{Lomb1976, Scargle1982, Horne1986}. For $N_0$ data points in the time series spanning a total length of $T$ in unit of days, we sampled the periodogram at $N_i$ independent frequencies \citep{Horne1986} from $1/T$ to $N_0/(2T)$ (which would be the Nyquist frequency if data were evenly sampled); the resolution of the periodogram is thus $\Delta f = (N_0/2-1)/(TN_i)$. Plotting power as a function of $f$ for all test frequencies, the dominant peak at frequency $f$ or period $P = 1/f$ then signals a significant variation at that frequency or period.

When identifying periodic sources from their periodogram peaks, we also took advantage of the redundancy of PS1 MDS monitoring in four filters ($g,r,i,z$), each with a slightly different observing cadence due to weather and technical downtime, to help filter out false detections by requiring periodogram peaks are coherently detected (within a 10\% error) in at least three filters. In each filter, the error of the peak due to noise can further be calculated as $\Delta f = 3\sigma_{r}/(4\sqrt{N_{0}}TA_{0})$  \citep{Horne1986, Kovacs1981} --- where we calculate $A_{0}$ as the best-fit sinusoidal amplitude of the light curve phase-folded on the averaged period $\bar{P}$ and $\sigma_{r}$ as the standard deviation of residuals after subtracting the signal from the light curve --- which gives us an error on the detected period: $\delta P = \delta f/f^2$. The total uncertainty of the detected period is calculated by putting the theoretical and measured errors in quadrature: $(\Delta P)^2 = (\sqrt{\sum{\delta P_{i}^2}}/4)^2+\sum{(P_{i}-\bar{P})^2}/(N-1)$ where $i$ = 1...N is the index of the coherent filter.

\begin{figure*}[hb]
\centering
\epsfig{file=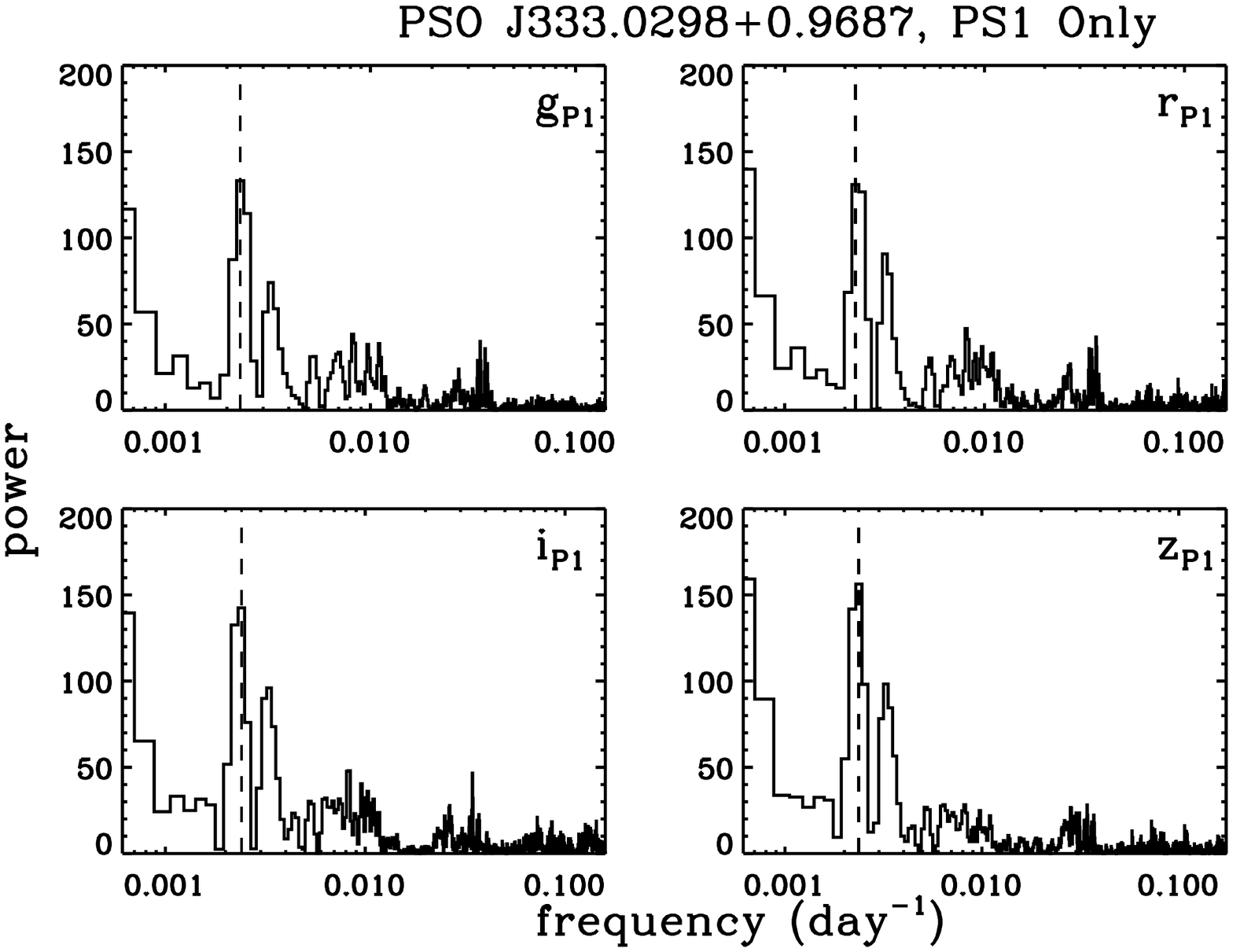,width=0.4\textwidth,clip=}
\epsfig{file=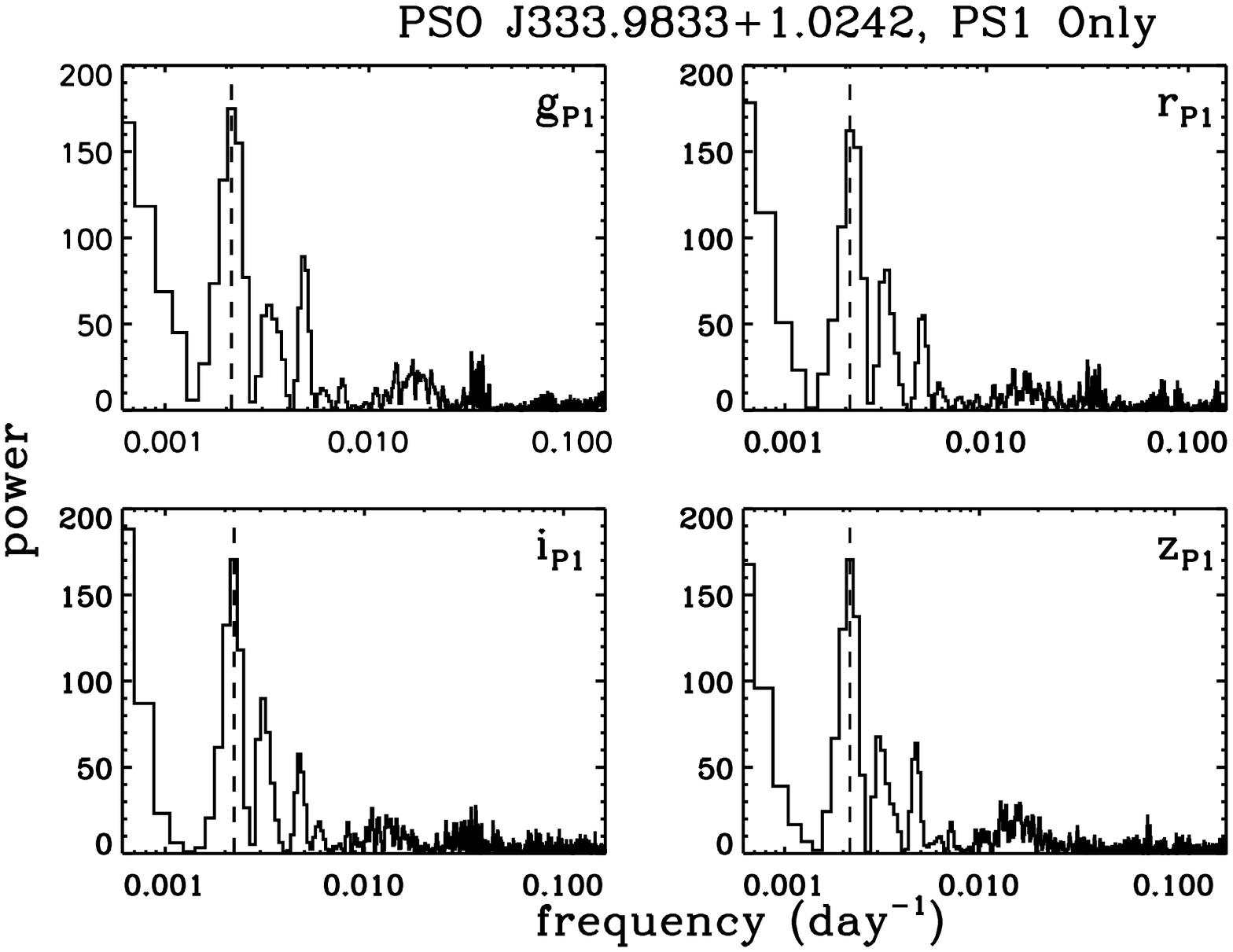,width=0.4\textwidth,clip=}
\epsfig{file=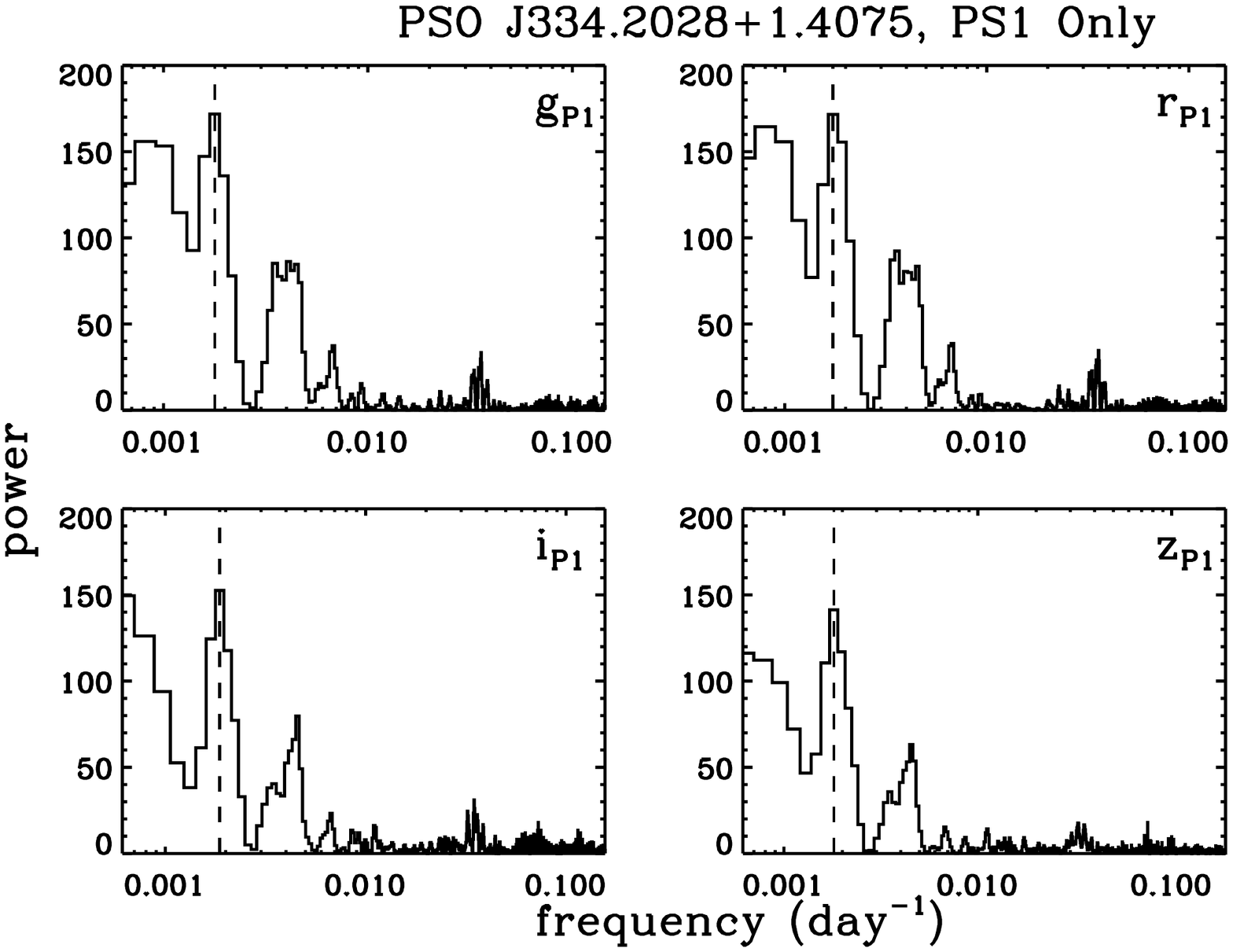,width=0.4\textwidth,clip=}
\caption{As part of our periodic quasar selection, we ran the Lomb--Scargle periodogram on PS1 light curves and selected the sources that have a coherent period detected in all four filters with high significant factors. In each set of panels, the coherent peak was marked with a dashed line in each filter.}
\label{fig:pgram}
\end{figure*}

We calculate the S/N ratio of the sinusoidal fit as $\xi=A_{0}^2/(2\sigma_{r}^2)$, where $\sigma_{r}$ is the standard deviation of the model-fit residuals \citep{Horne1986}. We mask any outliers that deviate from the mean by more than 4.5 $\sigma$ and require that candidates have $\xi>3$ in at least one filter. Finally we require at least 1.5 cycles of variation, in accordance with similar studies \citep{Graham2015, Charisi2016}; this limit on the maximum allowed period is also justified since spurious periods are oftentimes found on a timescale close to the total data length \citep{MacLeod2010}. Our selection leaves three candidate periodic quasars in MD09 (Table \ref{table:bythenumbers}). Their periodograms in $g_{\rm P1}$ $r_{\rm P1}$ $i_{\rm P1}$ $z_{\rm P1}$ are shown in Fig. \ref{fig:pgram}, and their complete PS1 light curves are presented in Figure \ref{fig:lc}. Note that in all three periodograms, variability power increases with lower frequency, which is a characteristic of red noise and a cause for concern in searching for periodicity. It is thus important to understand the false-alarm rate due to red noise and further test the sinusoidal model with extended baseline data (see discussion in \S\ref{sec:persistence}). 


\section{Extended Baseline Photometry}\label{sec:ext}

Historically, there have been claims of (quasi-) periodicity on a number of AGNs, but they failed to withstand re-analyses or follow-up observations (see e.g. review by \citealt{Vaughan2006QPOreview}). In the case of searching for light curve periodicity with a Fourier method, a finite temporal baseline makes the observer highly susceptible to ``red noise leak'' (see e.g. review by \citealt{Press1978} on red noise), where low frequency variations are transferred to the sampled high frequencies for objects with ``red'' power spectra of increasing power at low frequencies, such as AGNs and X-ray binaries.

Fortunately, all three of our candidates have extended baseline photometry, either from the archival database or our ongoing imaging campaign (Table \ref{table:extended}), giving us an advantage of testing the persistence of their periodic behavior by extending the baseline to $\sim2-3$ times the length of PS1 MDS with comparable photometric precision.

\begin{deluxetable}{lr}
\tablecaption{MD09 by the Numbers \label{table:bythenumbers}}
\tablehead{
\colhead{Category} & \colhead{Number}
}
\startdata
PS1 point sources & 40,488 \\
PS1$\times$CFHT quasars & 670 \\ 
PS1$\times$CFHT variable quasars & 104  \\
Variable quasars with coherent periodogram peaks & 77 \\
Candidates with $\xi>$3.0 in at least one filter & 6 \\ 
Candidates with $N_{\rm cycle}>$ 1.5 & 3 
\enddata
\end{deluxetable}


\begin{deluxetable*}{lll}
\tablecaption{PS1 Mean Magnitudes and Variability Amplitudes of Periodic Quasar Candidates \label{table:var_amp}}
\tablehead{
\colhead{PS1 Designation} & \colhead{$m$ (g,r,i,z)} & \colhead{$A_{0}$ (g,r,i,z)} 
}
\startdata
PSO J333.0298+0.9687 & (21.42, 20.94, 20.96, 20.95) & (0.68, 0.51, 0.53, 0.39) \\
PSO J333.9832+1.0242 & (18.97, 18.85, 18.79, 18.57) & (0.11, 0.10, 0.09, 0.07) \\
PSO J334.2028+1.4075 & (19.38, 19.28, 19.14, 18.94) & (0.13, 0.11, 0.08, 0.06) 
\enddata
\label{table:var_amp}
\end{deluxetable*}

\begin{deluxetable*}{llcc}
\tablecaption{Extended Baseline Photometry of Periodic Quasar Candidates \label{table:extended}}
\tablehead{
\colhead{PS1 Designation} & \colhead{Archival} & \colhead{Follow-up} & \colhead{UT Date of Follow-up Observations}
}
\startdata
PSO J333.0298+0.9687 & SDSS & \nodata & \nodata \\
PSO J333.9832+1.0242 & SDSS & \nodata & \nodata \\
PSO J334.2028+1.4075 & GALEX & DCT15Q2/Q3/16Q2/Q3 & 2015 May 28, 2015 Sep 17, 2015 Sep 19, 2016 May 15, 2016 July 10 
\enddata
\tablecomments{The column is empty if no follow-up imaging program is presented in the analysis.}
\end{deluxetable*}

\subsection{Follow-up Imaging}\label{sec:extended}

We have an on-going observing program at the Discovery Channel Telescope (DCT) in Happy Jack, Arizona, to further monitor candidate PSO J334.2028+1.4075 with its Large Monolithic Imager (LMI) in $g_{\rm SDSS}$ $r_{\rm SDSS}$ $i_{\rm SDSS}$ $z_{\rm SDSS}$ filters (Table \ref{table:extended}). Here we present data from four observing runs on UT 2015 May 28, 2015 September 17 and 19, 2016 May 15, and 2016 July 10.

Each observation had five exposures (taken in a dither pattern) in each filter on UT 2015 May 28 (5$\times$50s), UT 2015 September 17 ($g_{\rm SDSS}$ $r_{\rm SDSS}$ $i_{\rm SDSS}$, 5$\times$50s) and 19 ($z_{\rm SDSS}$, 5$\times$50s), UT 2016 May 15 (5$\times$100s), and UT 2016 July 10 (5$\times$100s). The images were reduced with standard \texttt{IRAF} routines, astrometry-corrected with \texttt{SCAMP} \citep{Bertin2006}, and co-added with \texttt{Swarp} \citep{Bertin2002}. For $z_{\rm SDSS}$ band images which are affected by fringe patterns, we constructed a master fringe map from all $z$ band images (with different telescope pointings) taken on one night using the \texttt{IDL} function \texttt{create\_fringes} \citep{Snodgrass2013}. Combining with a series of ``control pairs'' which mark the positions of adjacent bright and dark fringes in the map, we then subtracted a scaled fringe map from the image using the \texttt{IDL} function \texttt{remove\_fringes} \citep{Snodgrass2013}.

Using \texttt{SExtractor} \citep{Bertin1996}, we performed aperture photometry on the co-added image, with the aperture radius used in each filter being the typical full-width at half-maximum (FWHM) of the image and produced a catalog of detections in the LMI's 12'.3$\times$12'.3 field of view. We then cross-matched the catalog using a 1$\arcsec$ radius with all the point sources (\texttt{type} = ``star'') that are within a 6' radius from the target quasar with clean photometry (\texttt{clean} = 1) from the SDSS catalog. We excluded very bright objects ($m<16$ mag) to avoid saturated detections on the LMI images, and on the cloudy night (UT 20150917) and on all $z$ band images, we also constrained the fitting to the locus where $m<21$ mag. We iteratively removed outliers that systematically deviate from the residual fit by more than $0.2$ mag (for $m<22$ mag only) and fitted a linear function to the PSF magnitude \texttt{psfMag} vs. the \texttt{SExtractor} instrumental magnitude \texttt{mag\_aper} relation (we exclude the target quasar, which is variable, from fitting). Each residual plot was also visually inspected to confirm a good fit.

The magnitude error was calculated by taking the standard deviation of the residuals in the $\Delta m = 0.5$ mag vicinity of the target quasar. Finally, we converted the LMI photometric data to the PS1 magnitude for direct comparison with the light curves from MDS.

In their quasar variability study, \cite{Morganson2014} pointed out there are non-zero, albeit small, offsets for quasars after converting to PS1 magnitudes from the SDSS system. They adopt a third-order polynomial (derived for main sequence stars) to convert from SDSS to PS1 and add an additional average offset to correct for the color-dependent difference between the magnitudes. Since the \cite{Tonry2012photometry} filter transformations were also derived for stars (from synthetic magnitudes of stellar SEDs), we have the following options to correct between the SDSS and PS1 magnitudes in our light curves: (1) adopt the \cite{Tonry2012photometry} relation without any additional offset or correction; (2) adopt the \cite{Morganson2014} filter offsets;  (3) calculate redshift-dependent synthetic magnitudes (and thus offsets) from a composite quasar spectrum, where we redshift the composite quasar spectrum from \cite{VandenBerk2001} to the respective redshift of the candidate quasar and convolve it with the SDSS and PS1 filter sensitivity curves (airmass = 1.3 and 1.2, respectively) to calculate the synthetic magnitude in the respective bandpass and therefore the $m_{\rm P1}$-$m_{\rm SDSS}$ filter offset for each target. 

Even though we eventually adopted our redshift-dependent synthetic quasar correction as the most generic method, we note that the difference between the conversion equations are small ($\sim$ 0.01 mag), and, for quasars varying at the $>0.1$ mag level, as our candidates are, the different choices of filter conversion are unlikely to significantly change our conclusions with regard to the persistence of the variation.


\subsection{Pre-PS1 Archival Photometry}\label{sec:archival}

\begin{figure*}[h]
\centering
\epsfig{file=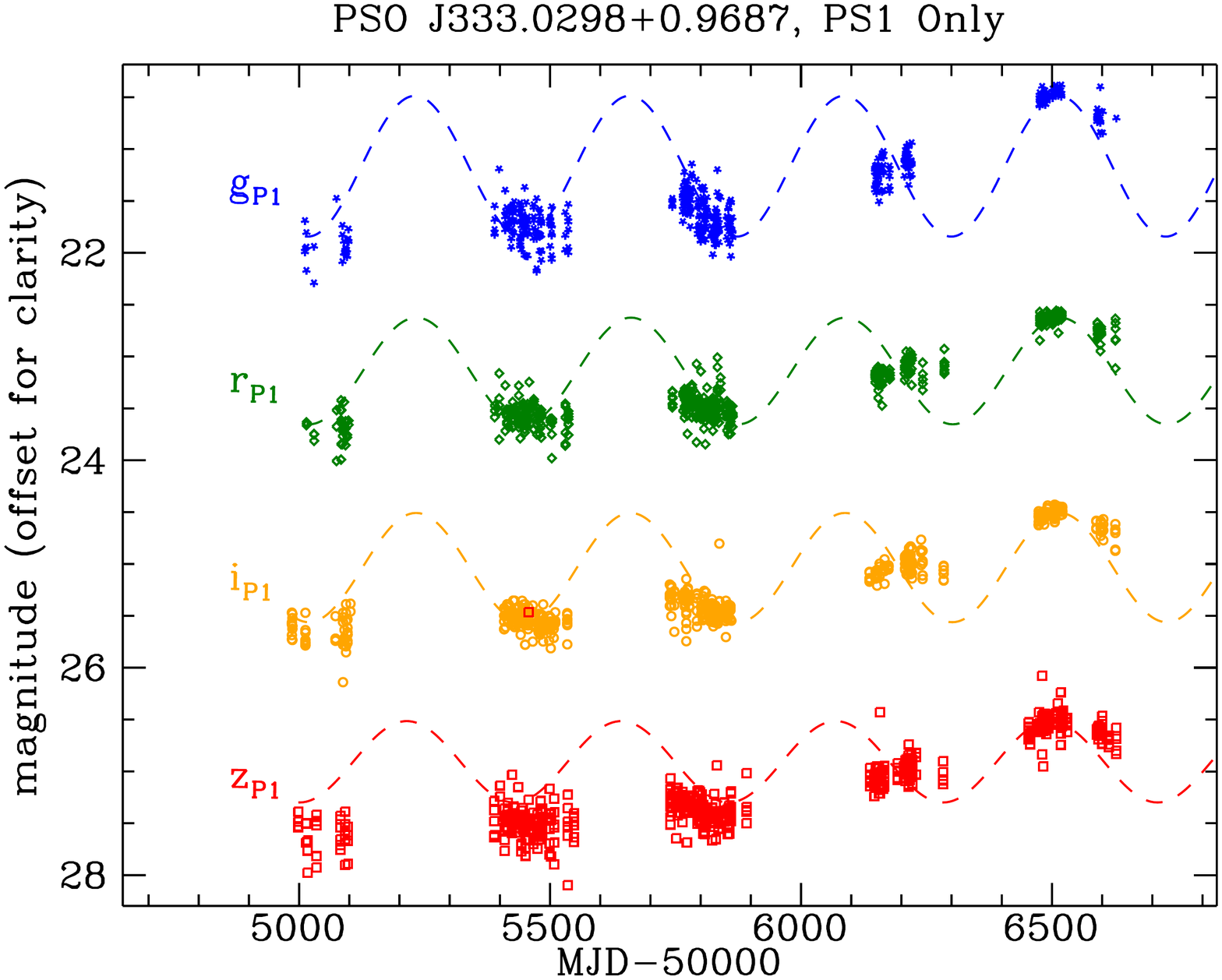,width=0.32\textwidth,clip=}
\epsfig{file=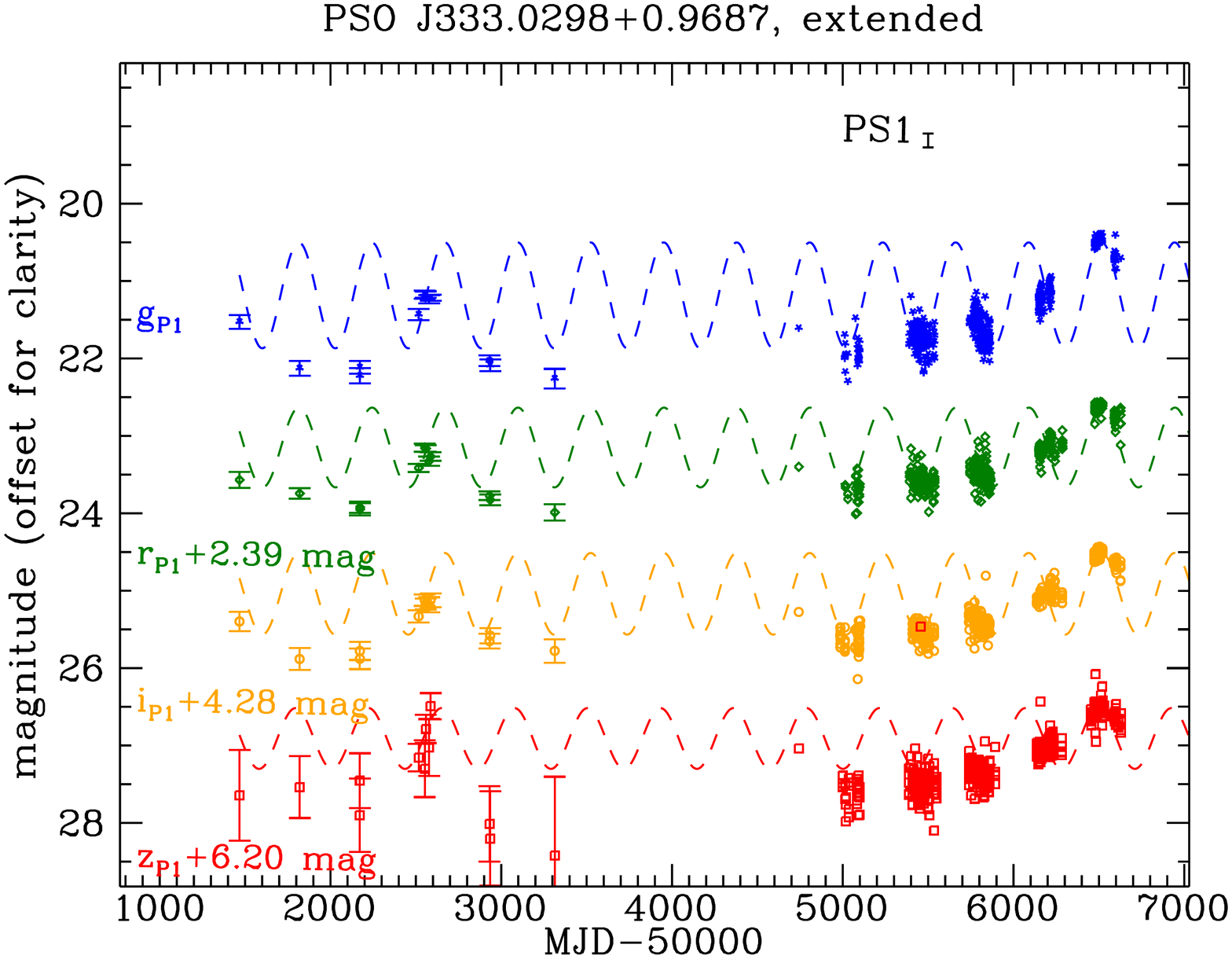,width=0.32\textwidth,clip=}
\epsfig{file=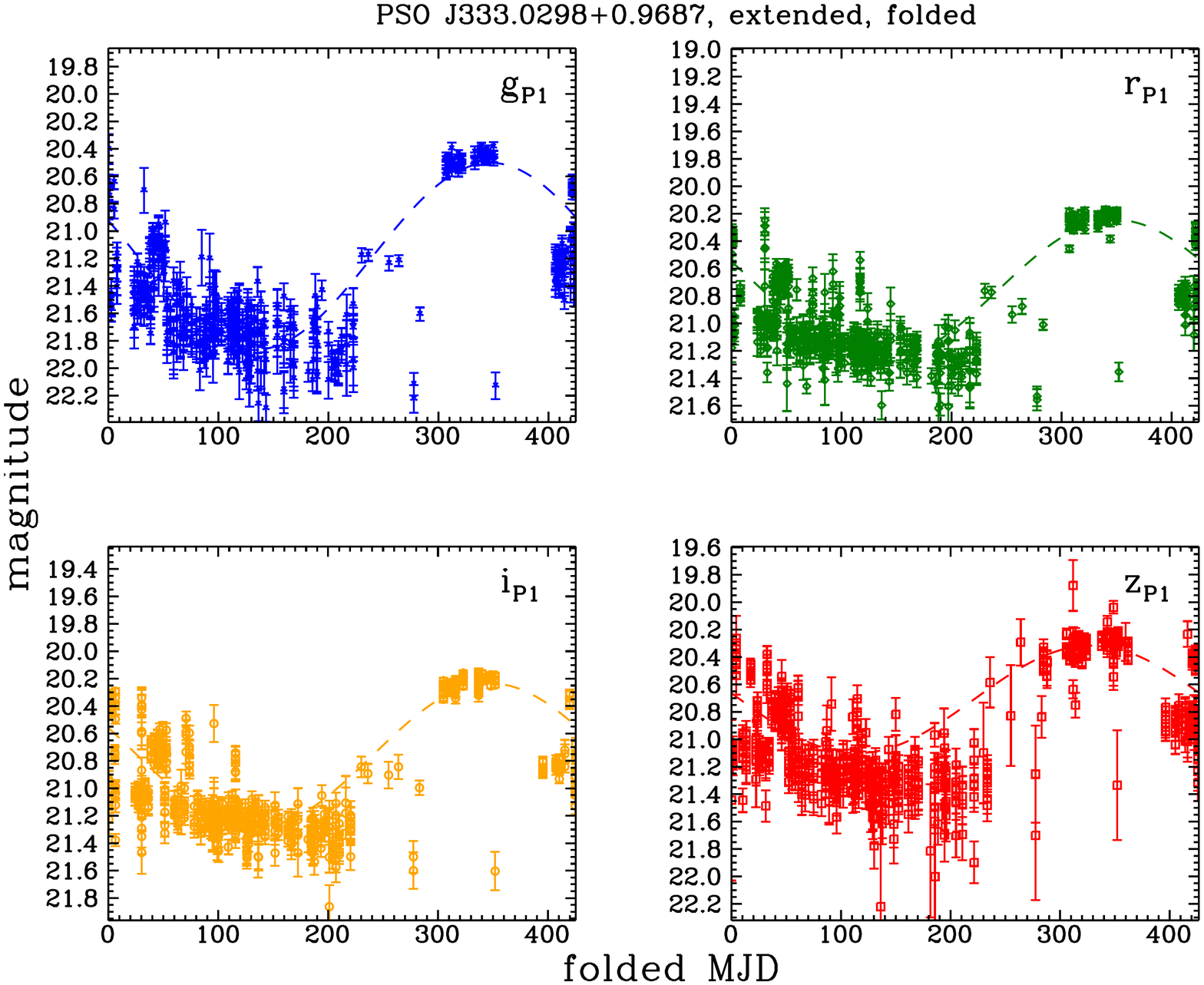,width=0.3\textwidth,clip=}

\epsfig{file=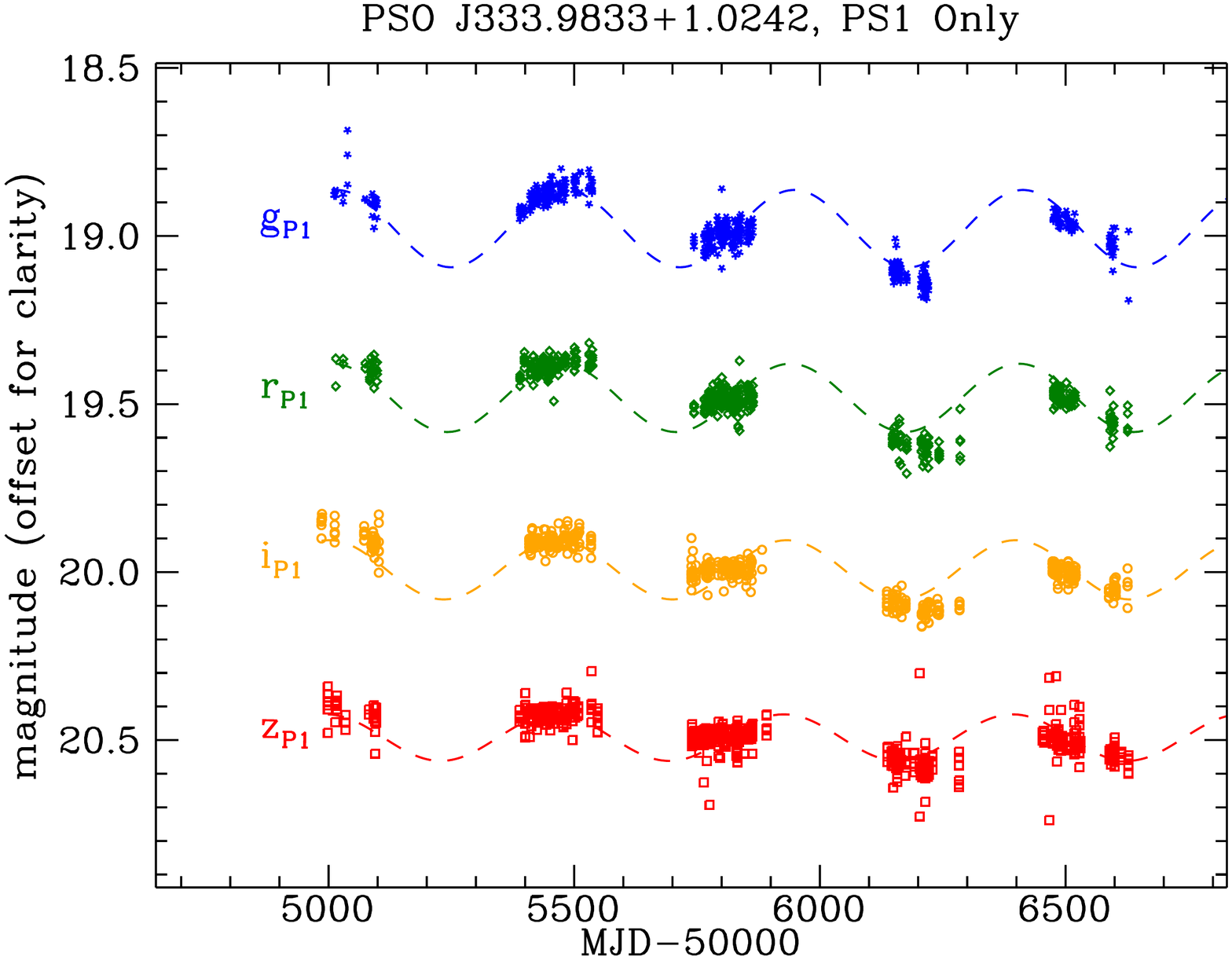,width=0.32\textwidth,clip=}
\epsfig{file=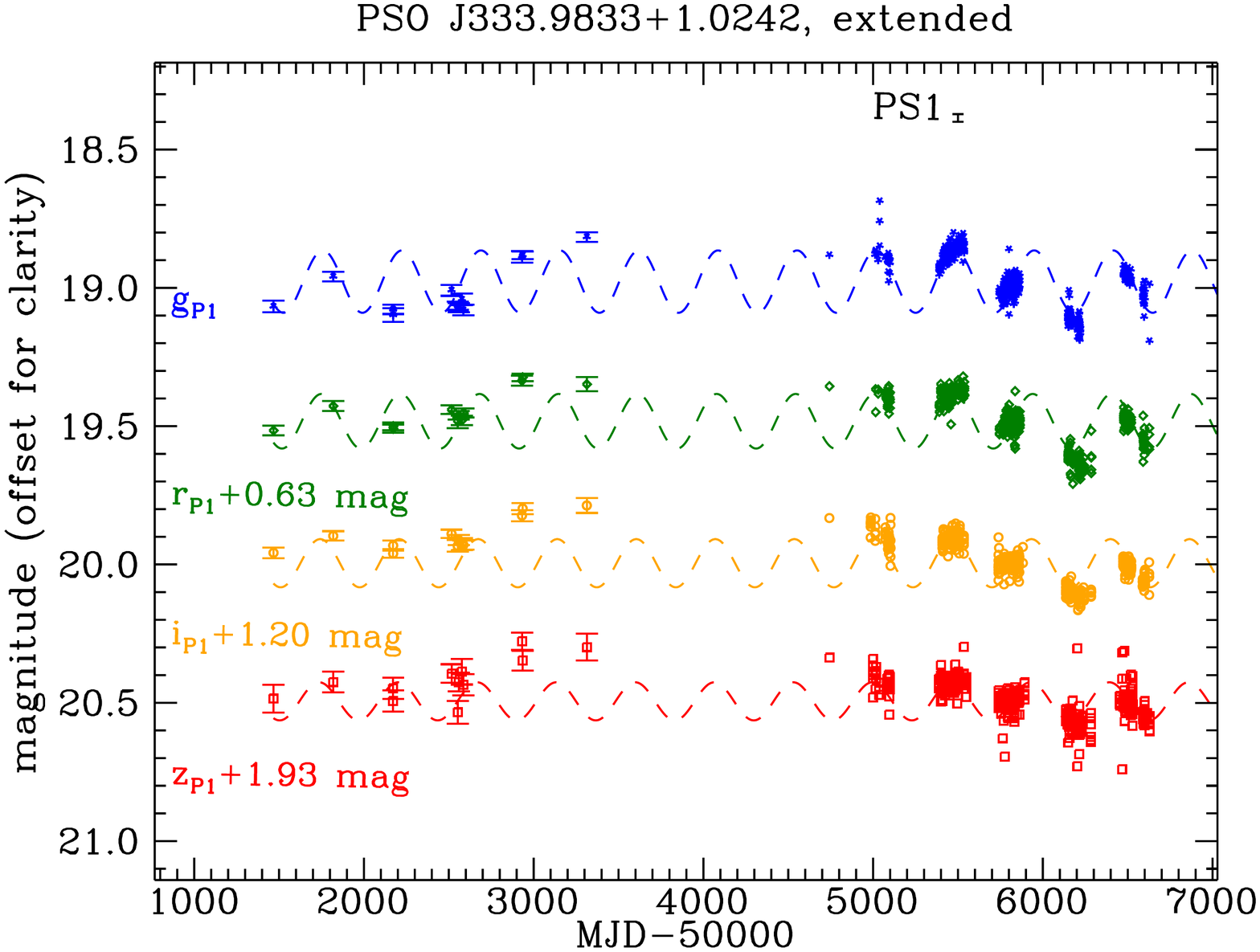,width=0.32\textwidth,clip=}
\epsfig{file=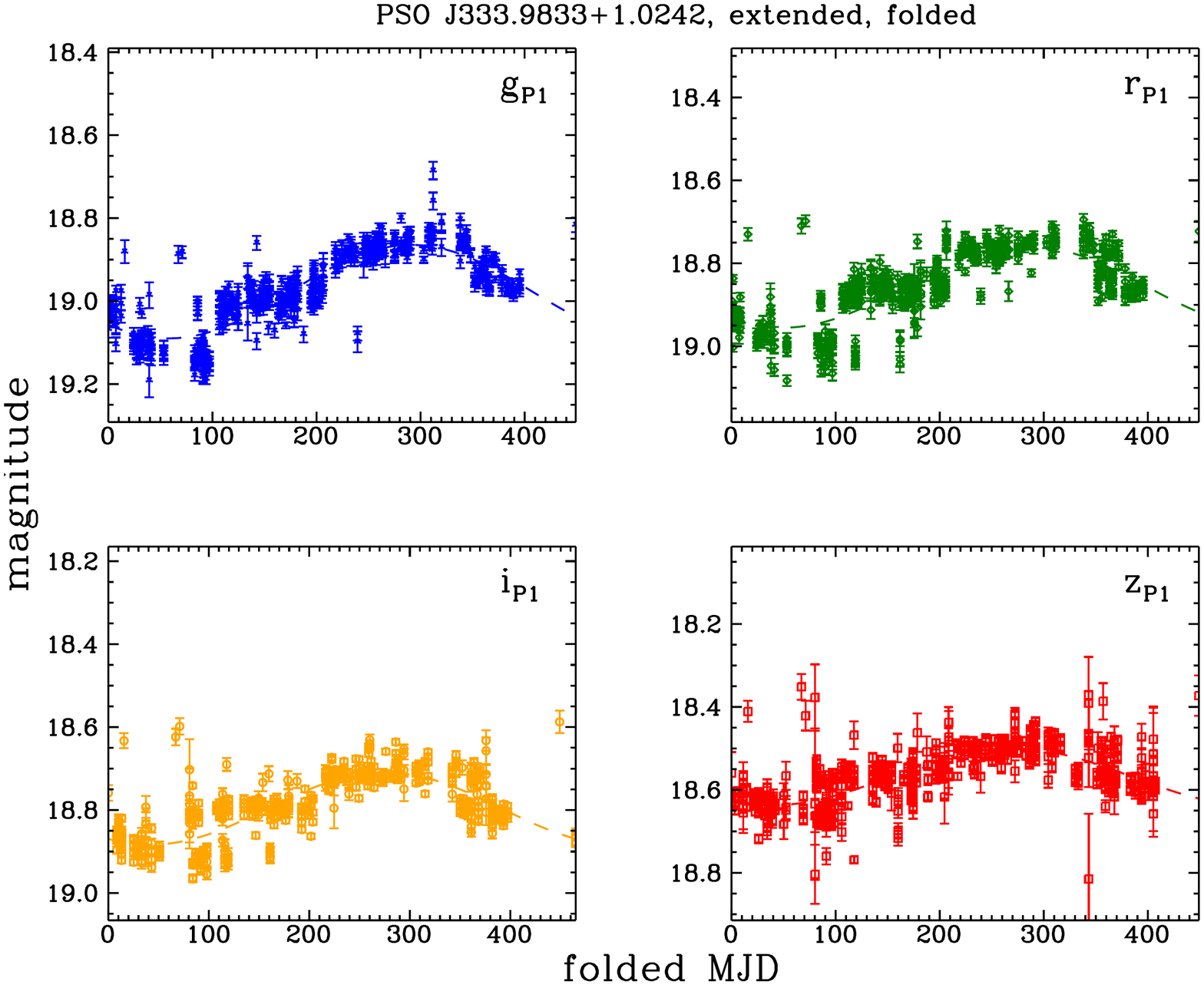,width=0.3\textwidth,clip=}

\epsfig{file=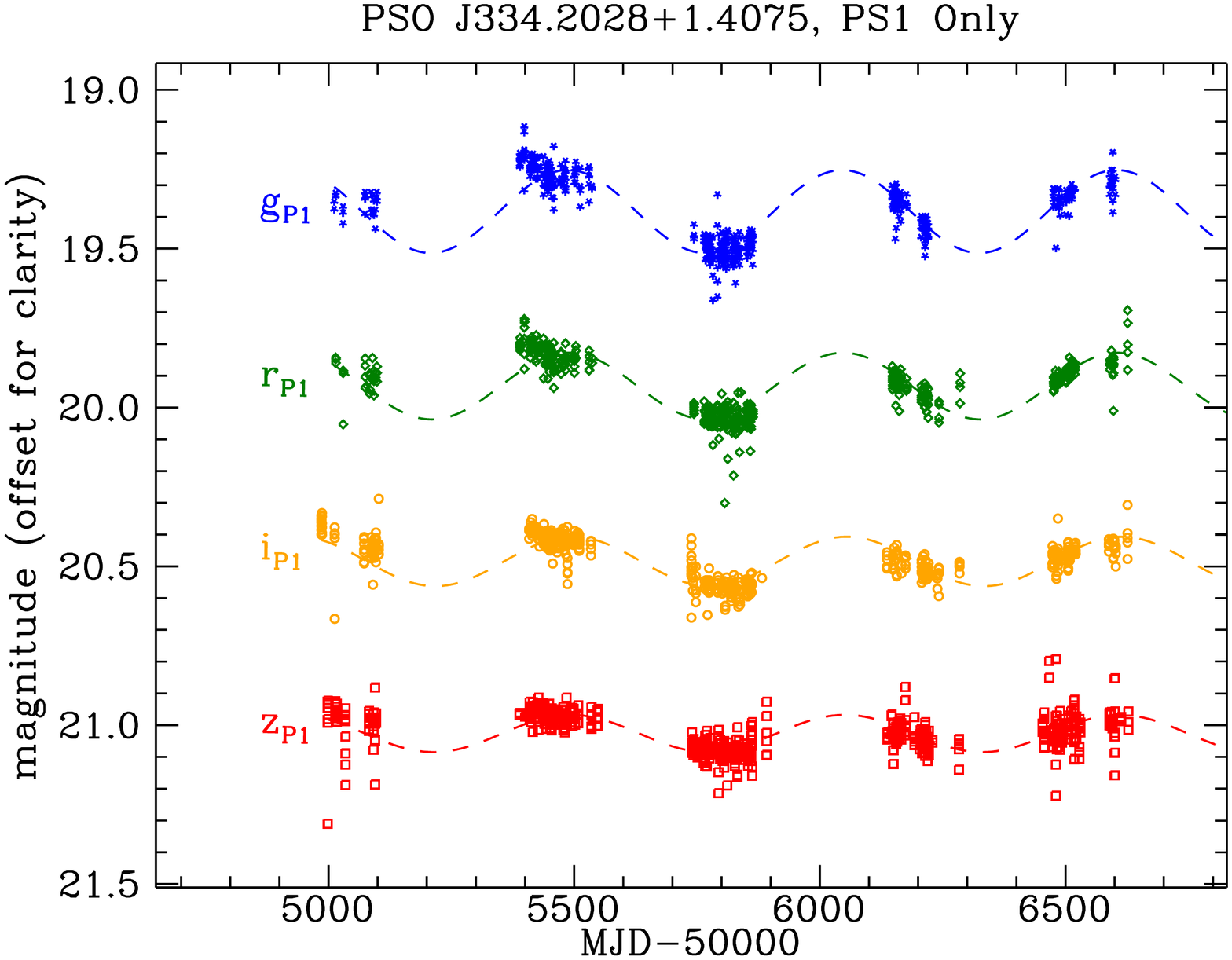,width=0.32\textwidth,clip=}
\epsfig{file=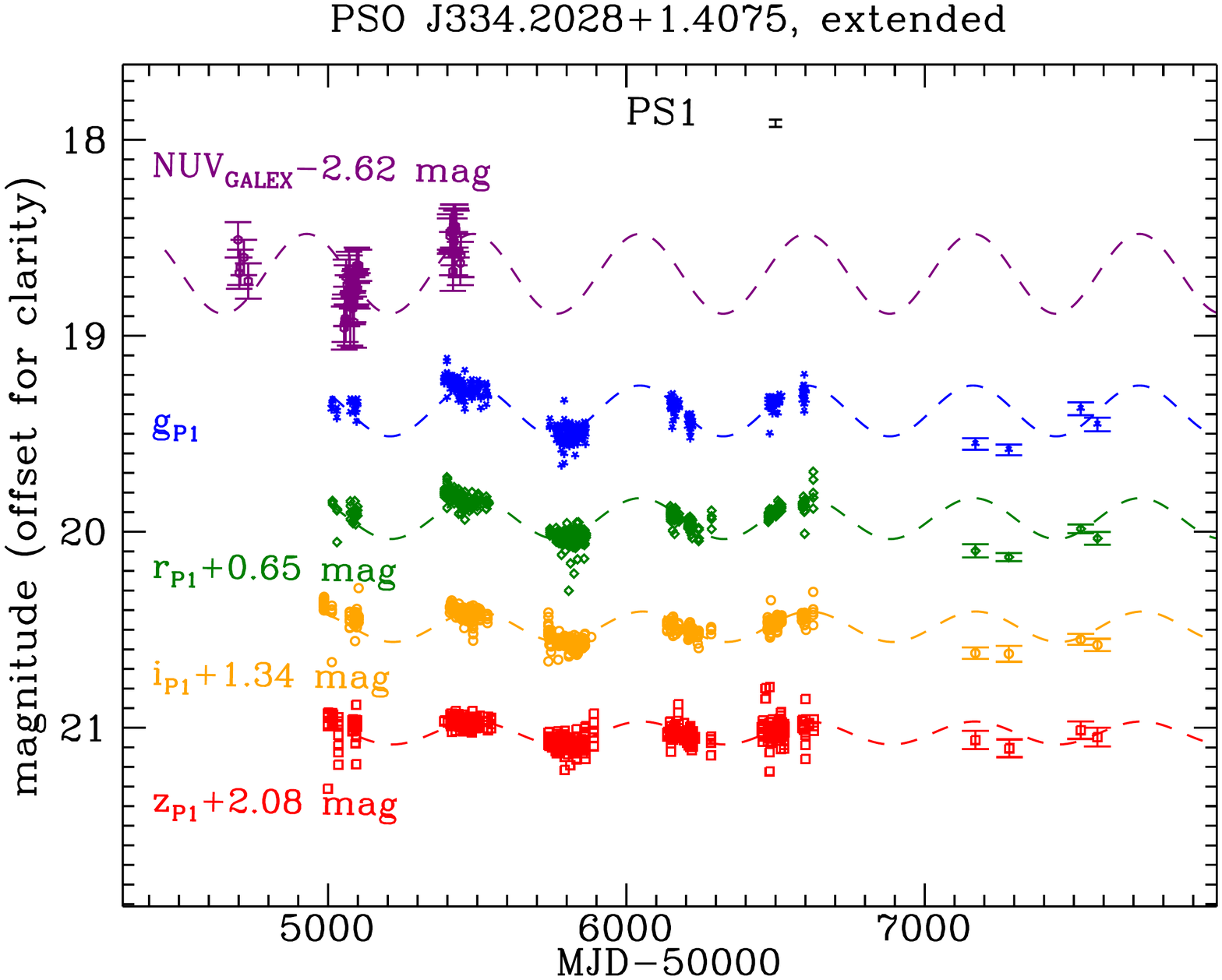,width=0.32\textwidth,clip=}
\epsfig{file=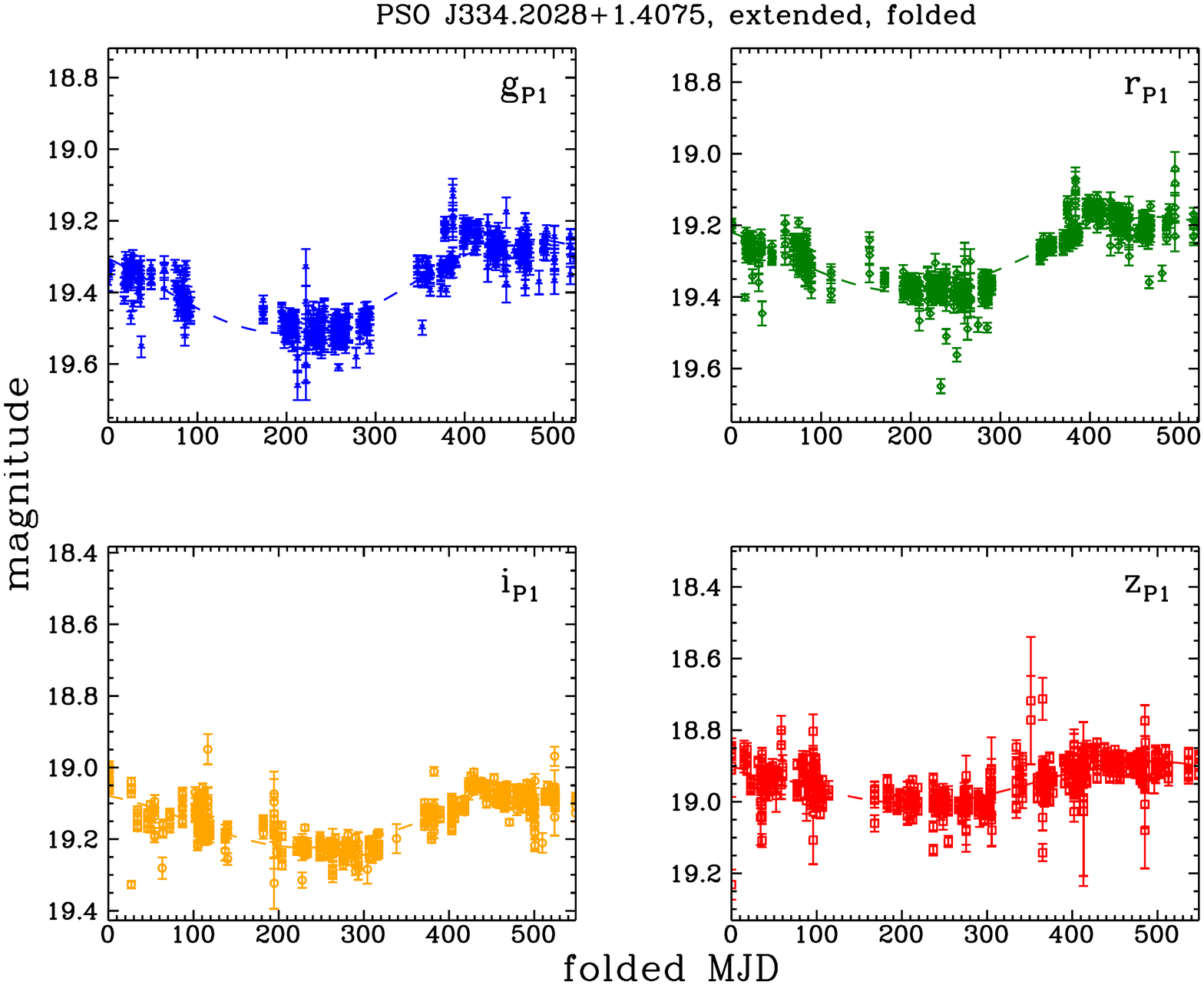,width=0.3\textwidth,clip=}
\caption{Left panels: PS1-only light curves in the $g_{\rm P1}$ $r_{\rm P1}$ $i_{\rm P1}$ $z_{\rm P1}$ filters. Light curves are offset for clarity. The light curves in different filters are fitted to sinusoidal functions of the same period ($\bar{P}$) and phase and of their respective best-fit amplitudes (dashed lines).  The PS1 photometric error bars are omitted for clarity; instead, the typical photometric error is indicated in the ``extended'' panel.
Middle panels: In the extended baseline light curves are fitted to sinusoidal functions of the same ``PS1 only'' period with the phase and amplitude being free parameters. S82 light curves and LMI data (taken in SDSS filters) have been converted to the PS1 photometric system. Light curves are also offset for clarity. For candidate PSO J334.2028+1.4075, its \textsl{Galaxy Evolution Explorer (GALEX)} UV light curve is also included, and we superimpose on it sinusoids of the scaled-up amplitude (purple) (see text for details).
Right panels: the extended baseline light curves are folded on the same period as the first two panels and fitted to sinusoidal functions.}
\label{fig:lc}
\end{figure*}

We retrieved pre-PS1 archival SDSS S82 PSF light curves in $g_{\rm SDSS}$ $r_{\rm SDSS}$ $i_{\rm SDSS}$ $z_{\rm SDSS}$ from SDSS-III DR12 \citep{Alam2015}.  The S82 magnitudes were converted to the PS1 system (\S\ref{sec:extended}) before being ``stitched'' to the PS1 light curves. The resulting PS1+SDSS light curves are shown in Figure \ref{fig:lc}.

Candidate PSO J334.2028+1.4075 also has a \textsl{Galaxy Evolution Explorer (GALEX)} Time Domain Survey \citep{Gezari2013} light curve available in the NUV band ($\lambda_{\rm eff}$ = 2316\AA) $\approx$ 1 year before the start of PS1 MDS. We superimpose on the $NUV$ light curve a sinusoid of the same period and phase as in the PS1 light curves and scale up the sinusoidal amplitude of the $g_{\rm P1}$ band ($\lambda_{\rm eff}$ = 4810\AA) by the observed exponential relation of variability amplitude as a function of (rest-frame) wavelength for quasars from \cite{VandenBerk2004} (see \S \ref{sec:bias}). The model is visually consistent with the larger variability amplitude of the NUV light curve (Figure \ref{fig:lc}). (The ordinate offset of the sinusoid is chosen such that it matches the mean magnitude of the NUV light curve.)

In explaining the observed periodic variability of the CRTS candidate PG 1302-102, \cite{DOrazio2015Nat} derived the expected variability amplitude ratio between the \textsl{GALEX} FUV and NUV and the CRTS $V$-band from spectral slopes, a corollary of their Doppler boosting model. However, we have shown, in addition to numerous previous studies (e.g. \citealt{VandenBerk2004, MacLeod2010, Gezari2013, Heinis2016AGN}) that a larger variability amplitude at shorter wavelengths is commonly observed in quasars and AGNs, and Doppler boosting is not unique in explaining the phenomenon.

\begin{deluxetable*}{lcccccc}
\tablecaption{SDSS to PS1 Filter Offsets from Synthetic Quasar Magnitudes \label{table:offset}}
\tablehead{
\colhead{PS1 Designation} & \colhead{$z$} & \colhead{$g_{\rm SDSS}$-$r_{\rm SDSS}$} & \colhead{$g_{\rm P1}$-$g_{\rm SDSS}$} & \colhead{$r_{\rm P1}$-$r_{\rm SDSS}$} & \colhead{$i_{\rm P1}$-$i_{\rm SDSS}$} & \colhead{$z_{\rm P1}$-$z_{\rm SDSS}$}
}
\startdata
PSO J333.0298+0.9687 & 1.284 & 0.2842 & -0.0197 & 0.0087 & -0.0005 & 0.0087 \\ 
PSO J333.9832+1.0242 & 2.234 & 0.1033 & 0.0052 & 0.0090 & -0.0108 & -0.0032 \\	 
PSO J334.2028+1.4075 & 2.070 & 0.1170 & -0.0127 & 0.0094 & -0.0060 & -0.0002  
\enddata
\tablecomments{$z$ is the spectroscopic redshift of the candidate. The SDSS colors $g$-$r$ and SDSS to PS1 magnitude offsets $m_{\rm P1}$-$m_{\rm SDSS}$ are calculated from synthetic magnitudes by convolving the composite quasar spectrum with the respective filter. The offset is then added to each LMI magnitude.}
\end{deluxetable*}

\begin{deluxetable*}{lrrcrr}
\tablecaption{Detected Period and Significance Factors of Periodogram-selected Candidates \label{table:P_xi}}
\tablehead{
\colhead{PS1 Designation} & \colhead{$\bar{P}\pm \Delta P$ (day)} & \colhead{$\xi$ (g,r,i,z)}  & \colhead{$N_{\rm cycle}$} & \colhead{$\xi$ (g,r,i,z)} & \colhead{$N_{\rm cycle}$} \\
& \colhead{(PS1 Only)} & \colhead{(PS1 Only)} & \colhead{(PS1 Only)} & \colhead{(extended)} & \colhead{(extended)}
}
\startdata
PSO J333.0298+0.9687 & 428$\pm$12 & (3.54, 2.82, 2.75, 1.09) & 3.8 & (2.96, 2.48, 2.42, 0.98) & 12.1 \\ 
PSO J333.9833+1.0242 & 465$\pm$11 & (3.93, 2.58, 2.16, 1.34) & 3.5 & (2.84, 1.99, 1.61, 1.07) & 11.1 \\ 
PSO J334.2028+1.4075 & 558$\pm$19 & (3.84, 2.74, 1.80, 0.91) & 2.8 & (3.33, 2.42, 1.66, 0.90) & 4.6 
\enddata
\tablecomments{The PS1+LMI light curve of PSO J334.2028+1.4075 totals a time span that corresponds to 4.6 cycles of sinusoidal variation (P = 558 d). The extended light curves span 5.2 cycles if the \textsl{GALEX} UV data are also taken into account.}
\end{deluxetable*}

\begin{deluxetable*}{lllcccccc}
\tablecaption{Spectroscopic Information and Inferred Binary Parameters of Periodic Quasar Candidates \label{table:spec}}
\tablehead{
\colhead{PS1 Designation} & \colhead{Spectroscopy} & \colhead{$f_{\lambda}$ (3000 \AA)} & \colhead{$\rm FWHM(Mg\,II)$} & \colhead{$\log{(M_{\rm BH}/M_{\odot})}$} & \colhead{$z$} & \colhead{$t_{\rm rest}$} & \colhead{$a$}  &  \colhead{$a$} \\
\colhead{} & \colhead{} & \colhead{$(\rm erg\,s^{-1}cm^{-2}\AA^{-1})$} & \colhead{$\rm (km\,s^{-1})$} & \colhead{} & \colhead{} & \colhead{$(\rm day)$} & \colhead{$(\rm pc)$} & \colhead{$\rm (R_{\rm s})$}
}
\startdata
PSO J333.0298+0.9687 & DCT15Q3 & $2.4\times10^{-17}$ & 8851 & 9.2$\pm$0.4 & 1.284 & 244.3 & 0.004 & 28 \\
PSO J333.9833+1.0242 & SDSS & $4.2\times10^{-17}$ & 6157 & 9.5$\pm$0.4 & 2.234 & 144.0 & 0.003 & 13 \\
PSO J334.2028+1.4075 & GS15A & $1.9\times10^{-17}$ & 5492 & 9.1$\pm$0.4 & 2.070 & 181.8 & 0.003 & 28 
\enddata
\tablecomments{Since the three candidates are no longer significant periodic candidates with the extended baseline, the binary model is \\ disfavored, and therefore we only show their binary parameters for comparison with other systematic searches.}
\end{deluxetable*}


\subsection{Testing the Persistence of Periodicity \\ with Extended Baseline Photometry}\label{sec:persistence}

We recalculated the S/N parameter $\xi$ for the extended light curves by forcing the same $\bar{P}$ detected in PS1 only light curves (Table \ref{table:P_xi}). Though all candidates still have high significance values ($\xi (g) \sim$ 3) and the extended data have variation amplitudes similar to their model sinusoidal amplitudes (Table \ref{table:var_amp}), the extended light curves do not agree with the extrapolation of their respective ``PS1 only'' sinusoidal models and the periodic oscillations are not persistent.

The three candidates were selected by first calculating their significance with respect to the null hypothesis of white noise (i.e. constant power over frequencies) (\S\ref{sec:periodic}). Previous systematic searches also assumed the null hypothesis of damped random walk noise (DRW, \citealt{Kelly2009}) to calculate the false-alarm rate and thus statistical significance of their selected binary candidates (\citealt{Graham2015}; L15; \citealt{Charisi2016}). (However,  we note that the extended baseline data in \cite{Charisi2016} show that their DRW simulations underestimate the false-alarm rate.) The DRW null hypothesis is motivated by results from quasar light curve analyses which demonstrate that the DRW model is a good description of normal quasar variability \citep{Kelly2009, MacLeod2010}. The power spectrum of the DRW process is $P(f) = 2\sigma^2\tau^2/[1+(2\pi\tau f)^2]$ -- where $\sigma^2$ is the short-timescale variance and $\tau$ corresponds to the characteristic timescale --- it has a power law slope of $-2$ at high frequencies ($f >(2\pi\tau)^{-1}$) and flattens to $0$ at low frequencies, analogous to the X-ray power spectrum of AGNs and X-ray binaries in the ``low hard'' state (e.g. review by \citealt{McHardy2010}, pp 203-32). 

However, regardless of the model chosen for the power spectrum of quasar variability to evaluate the significance of the period detection, we are fundamentally limited by the several-year temporal baseline of current time domain surveys. Vaughan et al. (2016) show that mock light curves generated from both DRW and a broken power law power spectrum cannot be distinguished from a periodic signal over $\sim 2$ cycles, especially when adding photometric noise and down-sampling the light curve to the actual observing cadence. Fortunately, a periodic candidate can be favored or disfavored by observing it for a longer period of time (for a total of $\gtrsim$5 cycles, ideally with better sampling and photometric precision), as \cite{Vaughan2016} have suggested and as we have demonstrated in this paper.


\section{Black Hole Mass Estimates and Inferred Binary Parameters}\label{sec:mass}

In order to measure the total black hole mass of the system ($M_{\rm BH}$) and derive parameters under the binary model, we extracted the archival SDSS spectrum of candidate PSO J333.9833+1.0242. We were also able to acquire spectroscopic observations of the other two candidates from DCT or the Gemini-South Telescope: the spectrum of PSO J333.0298+0.9687 was obtained in 2015 Quarter 3 with DCT's DeVeny spectrograph with 300g mm$^{-1}$ grating and 1'' slit for an exposure time of 1400 s. The data were reduced with the standard \texttt{IRAF} routines. A Gemini GMOS-S long-slit spectrum was obtained for PSO J334.2028+1.4075 in the 2015A Semester (Program ID: GS-2015A-Q-17. PI: T. Liu) with R400 grating and 0''.75 slit for a total exposure time of 720 s. The Gemini spectrum was reduced with the Gemini \texttt{IRAF} package.

In both spectra acquired, we clearly captured the broad Mg II line, allowing us to use a combination of the broad line velocity and luminosity of the nearby continuum to estimate black hole mass of the system from \cite{McLure2004}:

\begin{equation}
\log\Big(\frac{M_{\rm BH}}{M_\odot}\Big) = 3.2\,\Big(\frac{\lambda L_{\lambda}}{10^{44}\,\rm ergs\,s^{-1}}\Big)^{0.62}\,\Big(\frac{\rm FWHM(Mg II)}{\rm km\,s^{-1}}\Big)^2.
\label{eqn:mgii}
\end{equation}

In order to measure the Mg II broad line width for the black hole mass estimate, we based our line fitting process on the prescription given by \cite{Vestergaard2001}, in order to subtract the iron pseudo-continuum emission that contaminates in the vicinity of the Mg II 2800\AA\ line: we first broadened the iron template presented in  \cite{Vestergaard2001} by convolving with a series of Gaussian kernels in an incremental step of 250 km s$^{-1}$, such that 1,000 km s$^{-1}$ $< \text{FWHM}_{\rm QSO} <$ 12,000 km s$^{-1}$. Then, in a fitting window of [2250,2650]\AA\ where iron emission is conspicuous \citep{Forster2001}, we compare the FWHM=2,000 km s$^{-1}$ template with the spectrum (from which a power law continuum was already subtracted) and iteratively determined a scale factor. We then compared the series of scaled and broadened templates with the spectrum to determine the best-fit $\text{FWHM}_{\rm QSO}$.

\begin{figure*}[ht]
\centering
\epsfig{file=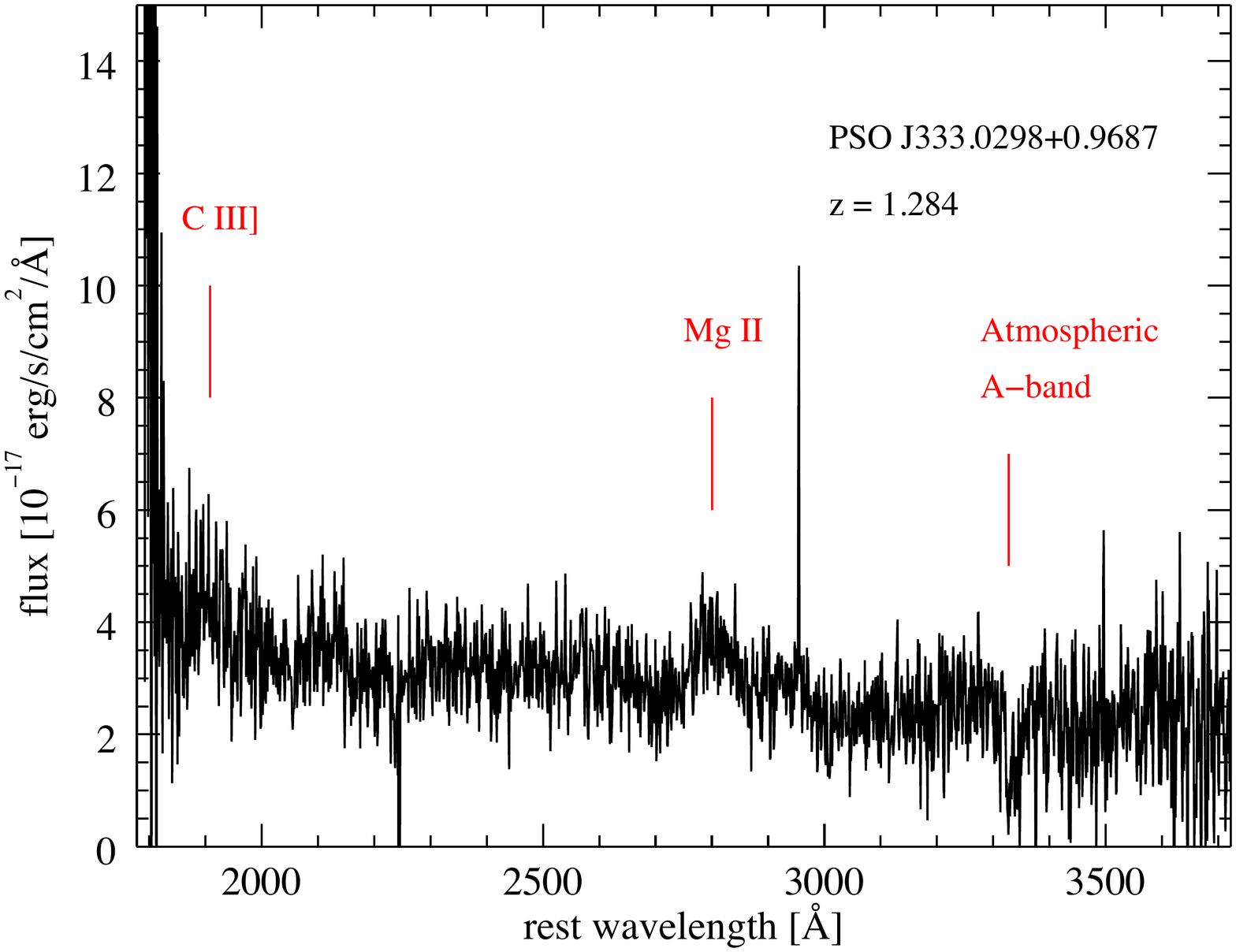,width=0.32\textwidth,clip=}
\epsfig{file=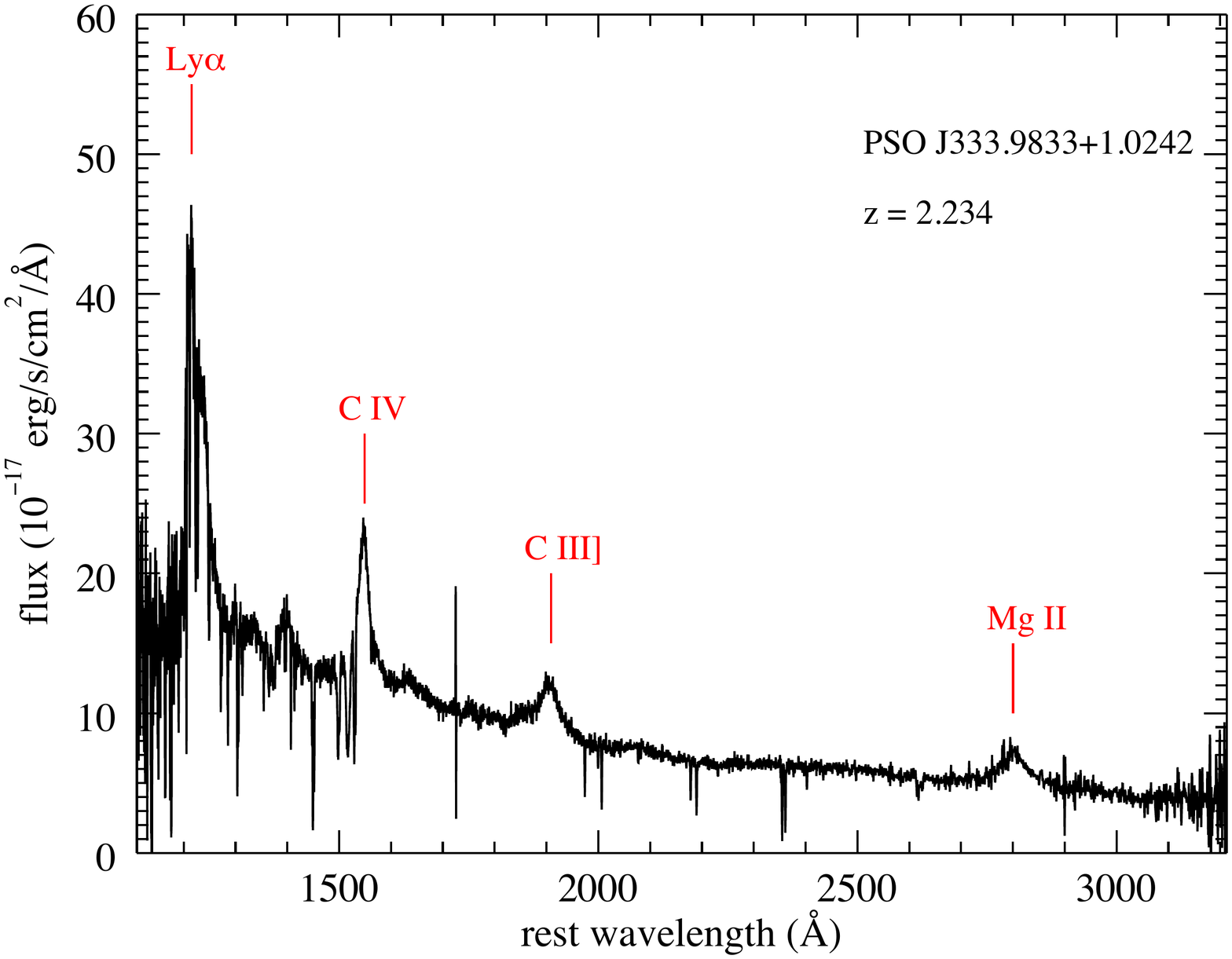,width=0.32\textwidth,clip=}
\epsfig{file=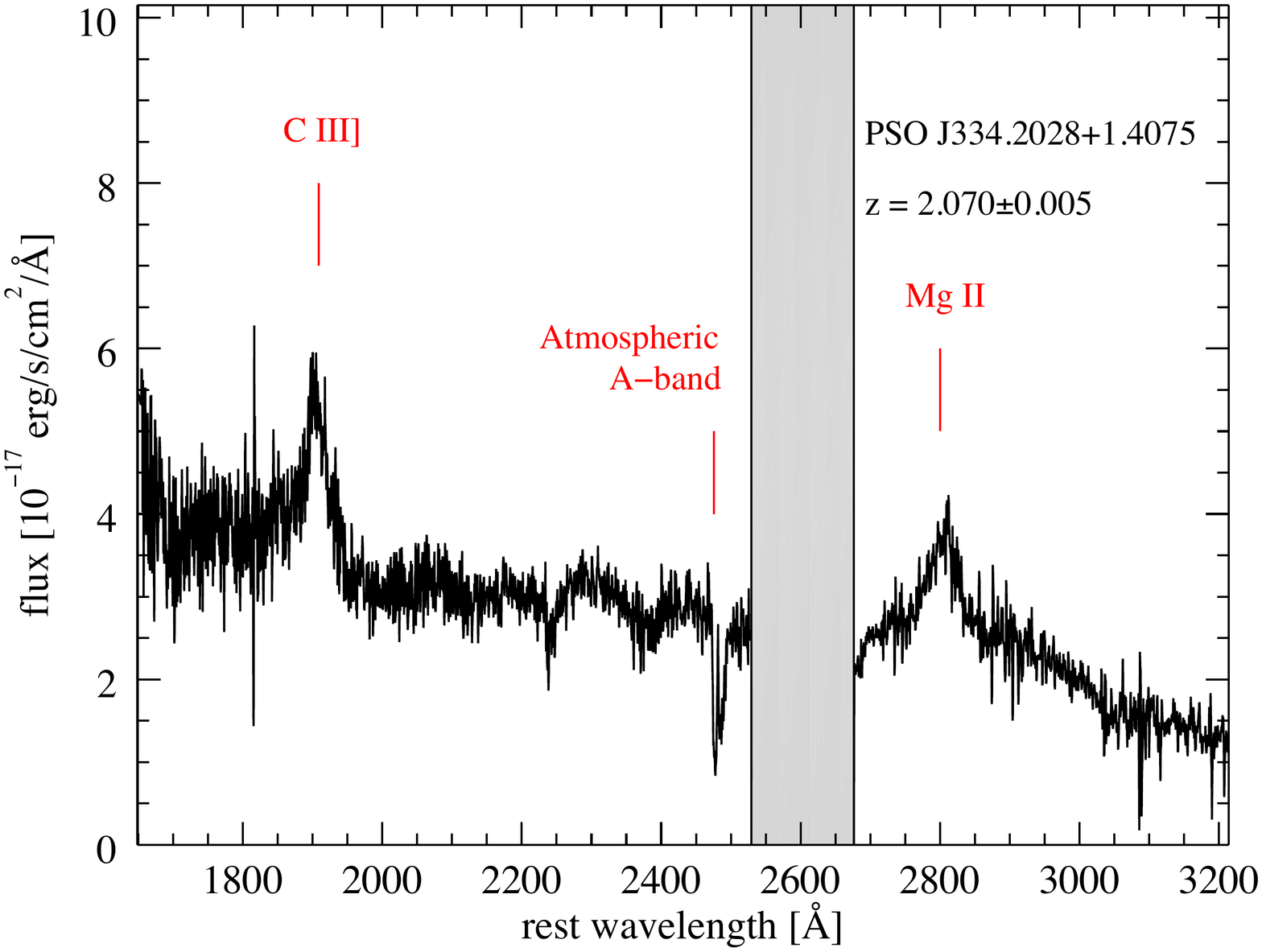,width=0.32\textwidth,clip=}
\epsfig{file=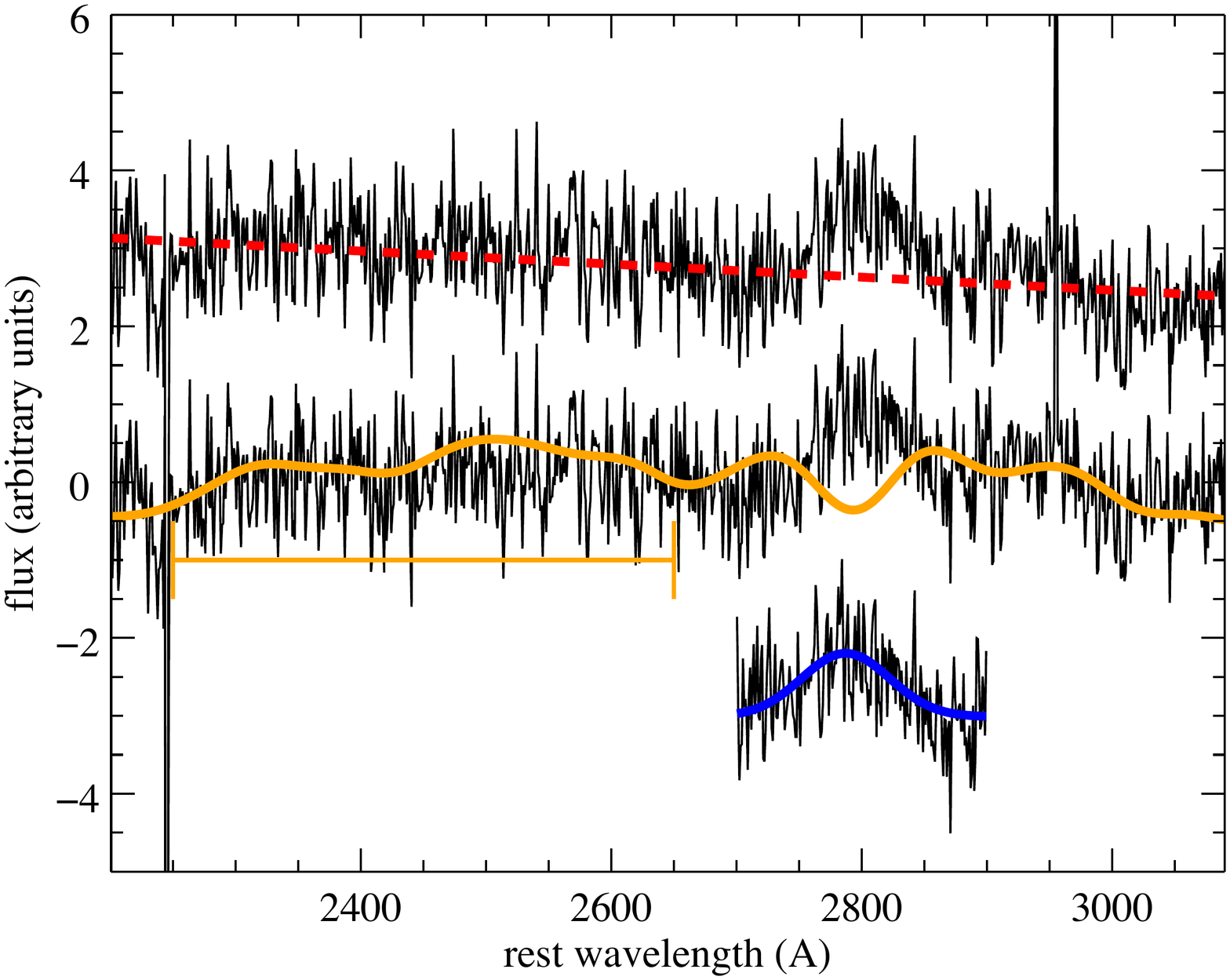,width=0.32\textwidth,clip=}
\epsfig{file=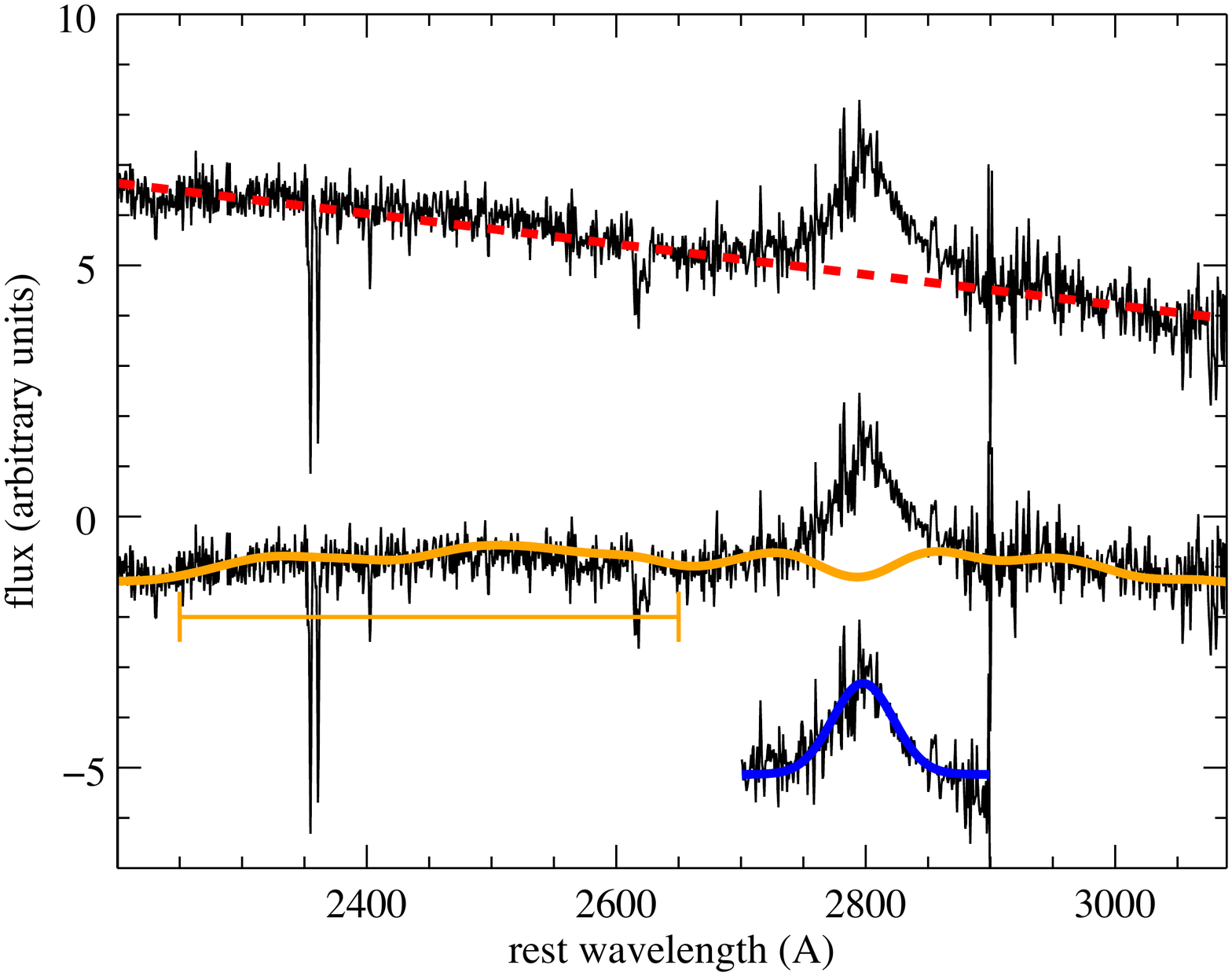,width=0.32\textwidth,clip=}
\epsfig{file=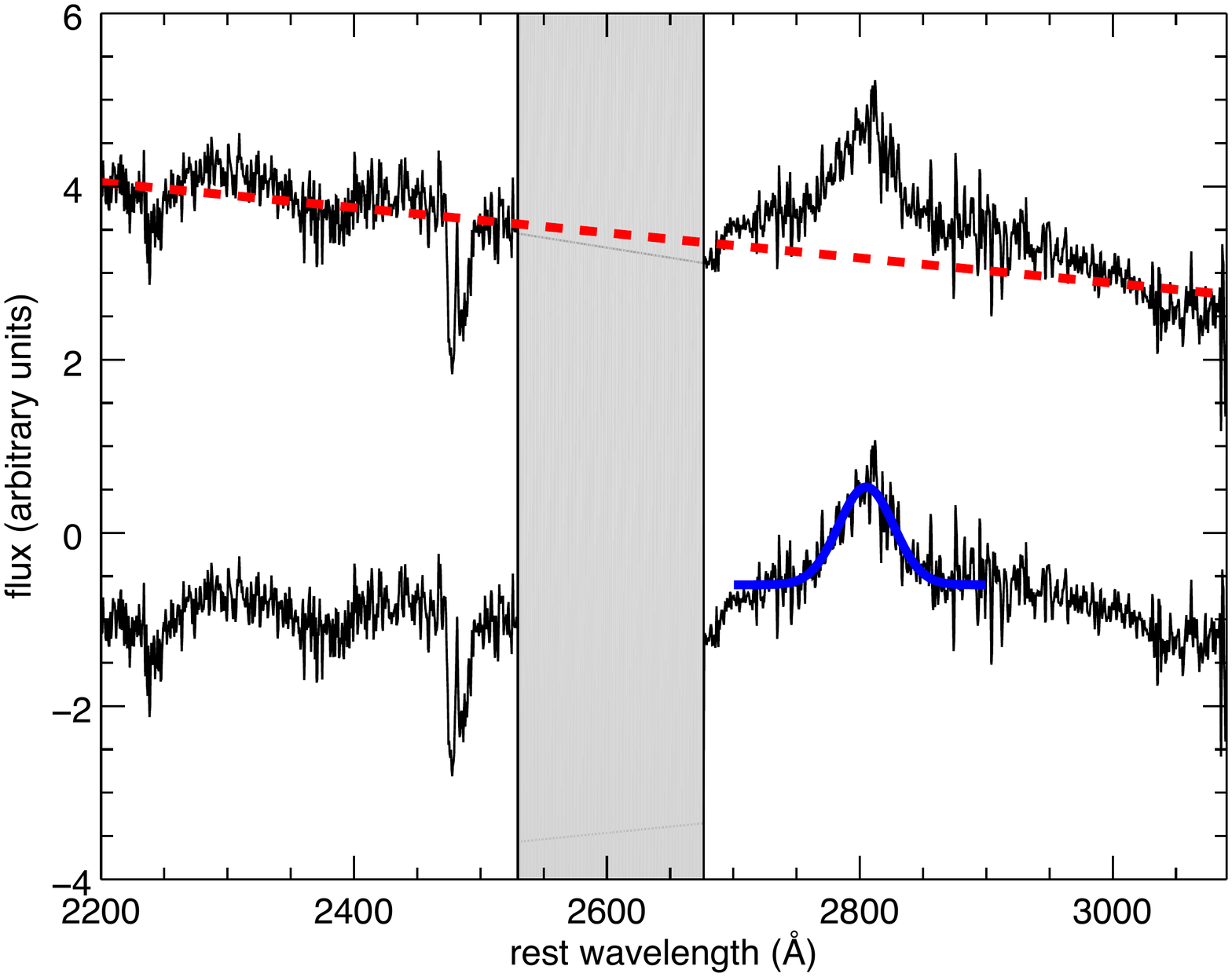,width=0.32\textwidth,clip=}
\caption{Top panels: MD09 candidate spectra from DCT/DeVeny, SDSS, and Gemini-South. Prominent emission features and the telluric absorption line at $7600-7630$ \AA\,(observed wavelength) are marked with red tick marks.
Bottom panels: we show the procedure by which we measure the Mg II line width: we fit and subtract the power law continuum (red dashed line) and iron emission in a spectral window (orange lines) and fit the Mg II line to a Gaussian (blue line). }
\label{fig:spec}
\end{figure*}

After fitting for the iron emission and subtracting from the spectrum, a Gaussian was then fitted to the Mg II broad line for the fitting range [2700,2900]\AA, whose FWHM was subsequently substituted into Equation \ref{eqn:mgii}. Any uncleaned sky lines were excluded from the fitting process, and the final continuum and iron-fitted spectrum was visually inspected to ensure the fitting is satisfactory (Fig. \ref{fig:spec}). Unfortunately, part of the Gemini spectrum (PSO J334.2028+1.4075) was affected by a misbehaving amplifier over the wavelength range where iron pseudo-continuum emission is strong. We had to mask the affected region and were not able to obtain a good fit of the iron emission; instead we only fit a power law continuum to the spectrum.

In the spectrum, we measured the continuum flux density at $\lambda = 3000$\AA\ ($f_{\lambda}$(3000 \AA)) from the continuum fitting and corrected for Galactic extinction using the dust map from \cite{SF2011} and the extinction curve from \cite{Cardelli1989}. With the redshift measured from the spectrum, we were then able to translate flux into luminosity $\lambda L_{\lambda} = \lambda 4 \pi D_{\rm L}^2 f_\lambda (1+z)$ and calculate the total black hole mass.

Though candidate PSO J333.9833+1.0242 has a measured black hole mass from \cite{Shen2008}, we applied the same line fitting and mass measurement routine to its SDSS spectrum to obtain a self-consistent measurement. We estimate $\log{(M_{\rm BH}/M_\odot)} = 9.5\pm0.4$, consistent with the \cite{Shen2008} mass of $\log{(M_{\rm BH}/M_\odot)} \approx 9.8$.

For PSO J334.2028+1.4075, we measured black hole mass $\log{(M_{\rm BH}/M_\odot)} = 9.1$ (with an error of $0.4$ dex associated with the black hole mass estimator Mg II; \citealt{McLure2002}); it is lower than $\log{(M_{\rm BH}/M_\odot)} = 9.97\pm0.5$ quoted in L15 --- which was estimated from C IV \citep{Vestergaard2006} and was not measured from an electronic spectrum --- but consistent with the previous black hole mass, considering the large scatter between the Mg II and C IV-based methods ($\log{(M_{\rm Mg II}/M_{\rm C IV})} = -0.06$ dex, with a dispersion of 0.34 dex; \citealt{Shen2008}).

Having obtained the black hole masses, we convert the observed variability period to the rest frame of the presumed binary: $t_{\rm var}$ = $t_{\rm orb} (1+z)$ and directly calculate binary orbital separation from Kepler's third law for a circular binary orbit:

\begin{equation*}
\frac{t_{\rm orb}^2}{a^3} = \frac{4\pi^2}{GM}\quad,
\label{eqn:kepler}
\end{equation*}

\noindent where $t_{\rm orb}$ is the rest-frame binary orbital period, $a$ is the orbital separation, and $M$ is the total mass of the system. The candidates' continuum flux density, Mg II line width, black hole mass, redshift, rest frame variability period $t_{\rm rest}$, and inferred binary orbital separation $a$ are tabulated in Table \ref{table:spec}. 

\begin{figure*}[h]
\centering
\epsfig{file=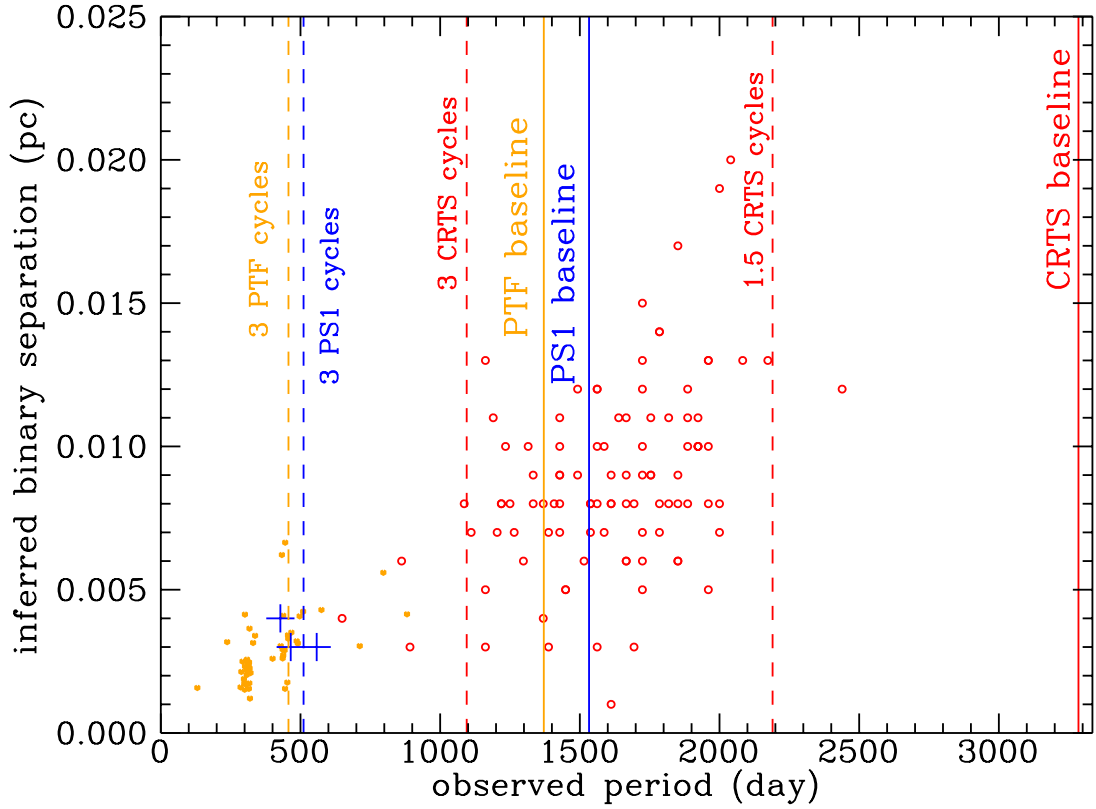,width=0.9\textwidth,clip=}
\caption{The blue plus signs mark our three periodic quasar candidates in PS1 MD09, which we classify as false alarms after failing the test of persistence over an extended baseline. The blue lines mark, from left to right, the length that corresponds to 3 cycles of variation over the MDS baseline and the 4.2-year MDS baseline. The 111 candidates from a systematic search in CRTS \citep{Graham2015} are in red circles. Red lines represent 3 CRTS cycles, 1.5 CRTS cycles (the minimum number of cycles required in their search), and the CRTS survey baseline (assuming the typical length of 9 years). The 50 candidates from PTF \citep{Charisi2016} are in orange dots. Orange lines mark 3 PTF cycles and the PTF baseline (assuming the typical length of 3.8 years).}
\label{fig:sep_tvar}
\end{figure*}

Even though the three periodic quasar candidates from our search in PS1 MD09 have been disfavored by our extended baseline analysis, we compare their observed period and inferred binary separation with search results from two other time domain surveys in Figure \ref{fig:sep_tvar}: CRTS \citep{Graham2015} and PTF \citep{Charisi2016}. Assuming the typical CRTS baseline of 9 years, all but seven of the 111 candidates claimed by \cite{Graham2015} have variations of less than 3 cycles (they require a minimum number of 1.5 cycles in their search), an insufficient data length for a robust periodicity detection according to \cite{Vaughan2016} (see our discussion in \S\ref{sec:persistence}).  As for the 50 candidates from PTF \citep{Charisi2016}, although the majority (82\%) of the candidates have more than 3 cycles of variation, a large fraction of them have observed periods clustered around one year (42\% of their candidates have periods between $300-400$ days), a potential sign of the aliasing effect of periodograms due to seasonal sampling \citep{MacLeod2010}, even though their DRW simulations were down-sampled to the observing cadence to account for this effect.

\section{Summary and Conclusions}\label{sec:conclude}

Periodic variability in quasars on the timescales of months to years has been theoretically predicted as a signature of an SMBHB. Recent simulations show that in triaxial galaxies (e.g. \citealt{Vasiliev2015}), the ``final parsec problem'' (e.g. review by \citealt{Milosavljevic2003}) is no longer an insurmountable problem that stalls binary evolution at $a>$ 1 pc separations and that binaries can evolve into the GW-dominated regime ($a \lesssim 10^{-3}$ pc) within a few Gyrs. A systematic search for periodic quasars in a large synoptic survey thus provides a novel method to search for SMBHBs in the final phase of their evolution and can potentially yield GW sources in the nano-Hz frequency regime which is accessible to pulsar timing arrays (PTAs) including NANOGrav \citep{McLaughlin2013} and the Parkes Pulsar Timing Array \citep{Hobbs2013}.

Our systematic search in the Pan-STARRS1 (PS1) MD09 field resulted in three periodic quasar candidates, from an initial sample of $\sim$ 700 color-selected quasars, that are apparently periodic over the PS1 baseline of 4 years. We further tested the persistence of their periodicity with archival light curves from SDSS Stripe 82 and followed up with imaging with the DCT. Archival \textsl{GALEX} photometry also confirms a larger amplitude of variation at shorter wavelengths, consistent with previous quasar variability studies. These extended-baseline data with photometric precision comparable to that of PS1 disfavor a simple sinusoidal model for the three candidates over an extended baseline of $\sim$ 5 -- 12 cycles. This corresponds to a detection rate of $<$ 1 out of 670 quasars ($\lesssim1.5\times10^{-3}$), which is still compatible with the theoretically predicted sub-parsec binary quasar fraction of $\lesssim$ 10$^{-3}$ out to $z = 1$ from cosmological SMBH merger simulations \citep{Volonteri2009}.  The detection rate per area ($<$ 1 in 5 deg$^{2}$) is also in agreement with the theoretical prediction of 100 quasars per 1000 deg$^{2}$ of search area (or 0.5 periodic quasars in 5 deg$^{2}$) from \citet{Haiman2009} for a flux-limited survey of quasars with $m_i < 22.5$ mag.  Our ongoing search over all 10 PS1 MDS fields, together with using nightly stacked images in the future which are $\sim$ 1 mag deeper, should increase our sensitivity to true SMBHBs by a factor of 100, and yield tens of promising SMBHB candidates for extended baseline monitoring and multiwavelength studies.

In comparison to other SMBHB searches, we note that there are two binary candidates with double broad-line features from a sample of $\sim$ 17,500 SDSS quasars (or a detection rate of $\approx$ 10$^{-4}$) for $z < 0.7$ \citep{Boroson2009}, consistent with the predicted SMBHB rate of $\sim$10$^{-4}$ ($z < 0.7$) by \cite{Volonteri2009}. We also note that \cite{Graham2015} imply a similar detection rate to our study of 68/$\sim$75,000 $\sim$ 0.9$\times$10$^{-3}$ (for quasars $z < 1$), and \cite{Charisi2016} find a detection rate of $\approx$ 1.4$\times$10$^{-3}$ for $z<3$ (or 0.9$\times$10$^{-3}$ for the sub-sample that remained significant after their re-analysis with extended data); however, see our discussions in \S\ref{sec:persistence} and the relevant parts in \S\ref{sec:mass} on the robustness of those claimed candidates.

We have demonstrated the power of an extended baseline in testing periodic quasar candidates in surveys whose temporal baselines (covering only $1.5-4$ cycles) are susceptible to false detections from red noise characteristic of normal quasar variability.  Fortunately, for most of the periodic quasar candidates discovered in recent optical time domain surveys, continued monitoring over the next few years can robustly test the persistence of the periodicity over a necessary number of cycles ($> 5$) to filter out false alarms, and verify strong SMBHB candidates for direct detection in GWs by PTAs.


\acknowledgements

TL thanks Cole Miller for discussions of the statistical significance of periodic detections compared to red noise, M. Vestergaard for providing an electronic copy of the iron template, and the anonymous referee for their helpful comments. SG is funded in part by NSF AAG grant 1616566.

This research has made use of the VizieR catalogue access tool, CDS, Strasbourg, France.

These results made use of the Discovery Channel Telescope at Lowell Observatory. Lowell is a private, non-profit institution dedicated to astrophysical research and public appreciation of astronomy and operates the DCT in partnership with Boston University, the University of Maryland, the University of Toledo, Northern Arizona University, and Yale University.

LMI construction was supported by a grant AST-1005313 from the National Science Foundation. The upgrade of the DeVeny optical spectrograph has been funded by a generous grant from John and Ginger Giovale.

Based on observations obtained at the Gemini Observatory (acquired through the Gemini Science Archive and processed using the Gemini IRAF package), which is operated by the Association of Universities for Research in Astronomy, Inc., under a cooperative agreement with the NSF on behalf of the Gemini partnership: the National Science Foundation (United States), the National Research Council (Canada), CONICYT (Chile), Ministerio de Ciencia, Tecnolog\'{i}a e Innovaci\'{o}n Productiva (Argentina), and Minist\'{e}rio da Ci\^{e}ncia, Tecnologia e Inova\c{c}\~{a}o (Brazil).

The Pan-STARRS1 Surveys (PS1) have been made possible through contributions of the IfA, the University of Hawaii, the Pan-STARRS Project Office, the Max-Planck Society and its participating institutes, MPIA, Heidelberg and MPE, Garching, JHU, Durham University, the University of Edinburgh, QUB, the HarvardSmithsonian CfA, LCOGT Inc., the National Central University of Taiwan, STScI, NASA under Grant No. NNX08AR22G issued through the Planetary Science Division of the NASA Science Mission Directorate, NSF under Grant No. AST-1238877, the University of Maryland, and Eotvos Lorand University.

Funding for SDSS-III has been provided by the Alfred P. Sloan Foundation, the Participating Institutions, the National Science Foundation, and the U.S. Department of Energy Office of Science. The SDSS-III web site is http://www.sdss3.org/.

SDSS-III is managed by the Astrophysical Research Consortium for the Participating Institutions of the SDSS-III Collaboration including the University of Arizona, the Brazilian Participation Group, Brookhaven National Laboratory, Carnegie Mellon University, University of Florida, the French Participation Group, the German Participation Group, Harvard University, the Instituto de Astrofisica de Canarias, the Michigan State/Notre Dame/JINA Participation Group, Johns Hopkins University, Lawrence Berkeley National Laboratory, Max Planck Institute for Astrophysics, Max Planck Institute for Extraterrestrial Physics, New Mexico State University, New York University, Ohio State University, Pennsylvania State University, University of Portsmouth, Princeton University, the Spanish Participation Group, University of Tokyo, University of Utah, Vanderbilt University, University of Virginia, University of Washington, and Yale University.


\bibliographystyle{apj}
\bibliography{ms_resubmit}


\end{document}